\documentclass[twocolumn,twocolappendix,usenames,dvipsnames]{aastex63}
\usepackage{amsmath}

\shorttitle{Early Formation of MW-like Disks}
\shortauthors{Semenov et al.}
\graphicspath{{./}{figures/}}

\usepackage{etoolbox}
\makeatletter
\patchcmd{\NAT@citex}
  {\@citea\NAT@hyper@{\NAT@nmfmt{\NAT@nm}\NAT@date}}
  {\@citea\NAT@nmfmt{\NAT@nm}\NAT@hyper@{\NAT@date}}
  {}
  {}
\patchcmd{\NAT@citex}
  {\@citea\NAT@hyper@{%
     \NAT@nmfmt{\NAT@nm}%
     \hyper@natlinkbreak{\NAT@aysep\NAT@spacechar}{\@citeb\@extra@b@citeb}%
     \NAT@date}}
  {\@citea\NAT@nmfmt{\NAT@nm}%
   \NAT@aysep\NAT@spacechar%
   \NAT@hyper@{\NAT@date}}
  {}
  {}
\patchcmd{\NAT@citex}
  {\@citea\NAT@hyper@{%
     \NAT@nmfmt{\NAT@nm}%
     \hyper@natlinkbreak{\NAT@spacechar\NAT@@open\if*#1*\else#1\NAT@spacechar\fi}%
       {\@citeb\@extra@b@citeb}%
     \NAT@date}}
  {\@citea\NAT@nmfmt{\NAT@nm}%
   \NAT@spacechar\NAT@@open\if*#1*\else#1\NAT@spacechar\fi%
   \NAT@hyper@{\NAT@date}}
  {}
  {}
\makeatother

\def\Mvir{M_{\rm vir}}
\def\Rvir{R_{\rm vir}}

\def\vrot{v_{\rm rot}}

\def\vsall{(v_{\rm rot}/\sigma)_{\rm all}}
\def\vsyoung{(v_{\rm rot}/\sigma)_{\rm young}}

\def\SFR{\dot{M}_{\star}}
\def\tspinup{t_{\rm spin\text{-}up}}

\def\pc{{\rm \;pc}}
\def\kpc{{\rm \;kpc}}

\def\kms{{\rm \;km\;s^{-1}}}
\def\Msunyr{{\rm \;M_\odot\;yr^{-1}}}
\def\Msun{{\rm \;M_\odot}}

\def\cc{{\rm \;cm^{-3}}}

\newcommand{\newtext}{}

\defcitealias{bk22}{BK22}

\begin{document}

\title{Formation of Galactic Disks I: Why Did the Milky Way’s Disk Form Unusually Early?}

\author[0000-0002-6648-7136]{Vadim A. Semenov}
\altaffiliation{\href{mailto:vadim.semenov@cfa.harvard.edu}{vadim.semenov@cfa.harvard.edu} \\ NHFP Hubble Fellow}
\affiliation{Center for Astrophysics $|$ Harvard \& Smithsonian, 60 Garden St, Cambridge, MA 02138, USA}

\author[0000-0002-1590-8551]{Charlie Conroy}
\affiliation{Center for Astrophysics $|$ Harvard \& Smithsonian, 60 Garden St, Cambridge, MA 02138, USA}

\author[0000-0002-0572-8012]{Vedant Chandra}
\affiliation{Center for Astrophysics $|$ Harvard \& Smithsonian, 60 Garden St, Cambridge, MA 02138, USA}

\author[0000-0001-6950-1629]{Lars Hernquist}
\affiliation{Center for Astrophysics $|$ Harvard \& Smithsonian, 60 Garden St, Cambridge, MA 02138, USA}

\author[0000-0001-8421-5890]{Dylan Nelson}
\affiliation{Universitat Heidelberg, Zentrum f{\"u}r Astronomie, Institut f{\"u}r theoretische Astrophysik, Albert-Ueberle-Str. 2, D-69120 Heidelberg, Germany}

\begin{abstract}
Recent results from spectroscopic and astrometric surveys of nearby stars suggest that the stellar disk of our Milky Way (MW) was formed quite early, within the first few billion years of its evolution. Chemokinematic signatures of disk formation in cosmological zoom-in simulations appear to be in tension with these data, implying that MW-like disk formation is delayed in simulations. We investigate the formation of galactic disks using a representative sample of MW-like galaxies from the cosmological-volume simulation TNG50. We find that on average MW-mass disks indeed form later than the local data suggest. However, their formation time and metallicity exhibit a substantial scatter, such that $\sim$10\% of MW-mass galaxies form disks early, similar to the MW. Thus, although the MW is unusual, it is consistent with the overall population of MW-mass disk galaxies. The direct MW analogs assemble most of their mass early, $\gtrsim 10$ Gyr ago, and are not affected by destructive mergers after that. In addition, these galaxies form their disks during the early enrichment stage when the interstellar medium metallicity increases rapidly, with only $\sim$25\% of early-forming disks being as metal-poor as the MW was at the onset of disk formation, [Fe/H]~$\approx -1.0$. In contrast, most MW-mass galaxies either form disks from already enriched material or experience late destructive mergers that reset the signatures of galactic disk formation to later times and higher metallicities. Finally, we also show that earlier disk formation leads to more dominant rotationally supported stellar disks at redshift zero.
\end{abstract}

\keywords{Galaxy formation, Galaxy disks, Milky Way disk, Star formation, Magnetohydrodynamical simulations}

\section{Introduction}

Most of the stars in the present-day Universe form in galactic disks similar in mass to our Milky Way (MW). When and how such disks form remains one of the central questions of modern astrophysics.
It was long thought that MW-like disks formed recently, at $z\lesssim2$. This picture was motivated by the observed irregular morphologies, high velocity dispersions, and ubiquity of bright UV clumps in typical star-forming galaxies at redshifts $z>2$ \citep[e.g.,][see \citealt{conselice14} and \citealt{forsterschreiber-wuyts20} for reviews]{forsterschreiber06,forsterschreiber09,forsterschreiber11,genzel06,genzel08,elmegreen07,conselice08,guo15,guo18,livermore15,wisnioski15,simons17}.
From the theory side, the galactic environment during early galaxy assembly is also expected to be very different from that in the local Universe with a prevalence of active galaxy mergers, rapid intergalactic gas accretion via cold streams perturbing the galaxy, high gas densities and fractions leading to various disk instabilities, and highly variable star formation activity \citep[e.g.,][see \citealt{somerville-dave15} for a review]{keres05,brooks09,dekel-birnboim06,fire,fire2,dekel20,vintergatan1,gurvich22}.

Until recently, the main window into the early Universe was provided by the Hubble Space Telescope (HST). Being sensitive to the rest-frame UV, the HST observations at $z>2$ are biased to actively star-forming environments, making its view of galaxy population in the early Universe incomplete.
Over the past few years, however, the field has been transformed thanks to the advent of ALMA and JWST.
ALMA observations of the [CII] fine-structure line at 158 $\mu$m, which is one of the major gas coolants that make it possible to probe interstellar medium (ISM) kinematics at these high redshifts, revealed the abundance of massive dynamically cold galactic disks at high redshifts, up to $z \sim 6$ \citep[e.g.,][]{smit18,neeleman20,pensabene20,rizzo20,rizzo21,fraternali21,lelli21,tsukui-iguchi21,herreracamus22,posses23,romanoliveira23}.
After the successful launch of JWST, it is now also feasible to study the high-redshift Universe in rest-frame optical light. The first months of JWST observations revealed the surprising prevalence of galactic disks out to $z \sim 8$, corresponding to $\lesssim 1$ Gyr after the big bang \citep{ferreira22a,ferreira22b,jacobs22,naidu22,nelson22,robertson23}.
These high-redshift disks revealed by ALMA and JWST are massive and are likely progenitors of massive ellipticals in the centers of galaxy groups and clusters at the present day.

Major insights into the formation of our own MW have recently been gained from the chemistry and kinematics of nearby stars in large spectroscopic and astrometric surveys like APOGEE, LAMOST, Gaia, and H3\footnote{Hectochelle in the Halo at High Resolution} \citep{bk22,bk23,conroy22,rix22,xiang-rix22}. By using chemical tagging, stellar age estimates, and kinematics, this ``galactic archeology'' approach allows us to obtain insight into how our galaxy may have evolved over cosmic time. Despite the differences in sample selection and analysis, these different data sets consistently show a three-stage picture of the MW's disk assembly (compare, e.g., Figure~4 in \citealt{bk22}, Figure~3 in \citealt{conroy22}, and Figure~6 in \citealt{rix22}):
\begin{itemize}
    \item \emph{Protogalaxy.} The most metal-poor stars, [Fe/H] $\lesssim -1.5$, are arranged in a kinematically hot component dominated by random velocities with only weak or no net rotation (first introduced by \citealt{bk22} as ``Aurora'').\footnote{Note that these authors separate stars formed in situ based on the [Al/Fe] abundances, which can be reliably done only at [Fe/H] $> -1.5$. For that reason, the stars with lower metallicities are missing from the Aurora population identified by the authors, although in reality the distribution of [Fe/H] in that population extends to lower metallicities \citep[see, e.g., Figure~3 in][]{conroy22}.} This population is thought to reflect the remnants of the earliest phase of galaxy formation associated with disordered gas accretion and active galaxy mergers. 
    The observed elevated abundances of globular cluster chemical markers (N, Al, O, and Si) also suggest that this pre-disk stage is associated with highly clustered star formation compared to the later disk stage \citep{bk22,bk23}. 
    Although on average the galaxy growth at this stage is formally dominated by the in situ star formation \citep[e.g.,][]{oser10,rodriguez-gomez16,clauwens18,tacchella19,davison20}, the contribution from merging with similar-mass galaxies can be significant \citep[$\sim 20\%\text{--}30\%$ of the total stellar mass increase; see, e.g.,][and Section~\ref{sec:results:protogalaxy} below]{santistevan20,renaud21}, making even the notion of a clear single MW progenitor ambiguous. To avoid this ambiguity, \citet{conroy22} collectively denoted all this early star formation occurring in major separate galaxies before the coalescence as the protogalaxy.
    
    \item \emph{Spin-up of a thick disk.} At higher metallicities,  $-1.5 \lesssim \text{[Fe/H]} \lesssim -1$, the median rotational velocity quickly increases to $\vrot \sim 150\kms$; i.e., the MW's stellar disk acquires most of its final net rotation, $\sim 220\kms$. This {spin-up} occurs over a range of [Fe/H] of only $\sim 0.3$ dex implying rapid---on the timescale of $\sim 1$ Gyr---formation of what is now seen as the {thick disk}. The data suggest that, in the MW, the spin-up occurs very early, within the first 1--2 Gyr after the Big Bang \citep{bk22,conroy22}. 
    
    \item \emph{Settling down to a thin disk.} At later times, the thick disk settles down into a {thin disk} as a result of secular evolution \citep[e.g.,][]{bournaud09,bird13,bird21,grand16,gurvich22} and/or a merger event \citep[e.g.,][]{brook04,brook12,belokurov20,bonaca20,vintergatan1,conroy22}. 
\end{itemize}

In this paper, we focus on the first transition; i.e., the spin-up of the initial thick disk. 
Over the past decade, galaxy formation simulations have been useful for investigating disk formation \citep[e.g.,][]{brook12,bird13,bird21,fire,fire2,agertz-kravtsov15,agertz-kravtsov16,auriga,pillepich19,stern19,stern20,stern21,stern23,vintergatan1,gurvich22,hafen22,yu22,mccluskey23}. One important insight was that galactic disks in the early Universe are substantially different from the present-day ones: they are more turbulent, compact, dense, and gas-rich, leading to more vigorous evolution with bursty star formation histories. This appeared puzzling in view of ALMA discoveries of dynamically cold disks at $z \sim 3\text{--}6$ (see references above), but many recent galaxy simulations can in fact produce such cold disks very early \citep[e.g.,][]{kohandel20,kretschmer22,tamfal22}. However, there is still no consensus on how galactic disks form, especially in such an early Universe.

As \citet[][hereafter \citetalias{bk22}]{bk22} have shown, state-of-the-art cosmological zoom-in simulations of MW-like galaxies \emph{qualitatively} reproduce a sharp spin-up feature in stellar $\vrot$ vs.[Fe/H], but it occurs at metallicities 3--10 times higher than observed.\footnote{This range corresponds to the galaxy-to-galaxy variation of the metallicity at which median $\vrot$ reaches roughly half of its final value (see Figure~17 in \citetalias{bk22}).} Such an offset implies that, in simulations, the formation of the MW disk is delayed by several Gyrs. Interestingly, this result appears to be independent of the modeling of star formation and feedback: \citetalias{bk22} find a similar shift in simulations with both explicitly modeled cold ISM formation and feedback injection \citep[FIRE-2;][]{fire2,fire2-release} and with an effective equation-of-state modeling \citep[Auriga;][]{auriga,aurigaia}. At the same time, some tailored simulations of local group analogs appear to produce the spin-up feature at low metallicities, in better agreement with the local data, even though there is still a significant galaxy-to-galaxy scatter \citep[see Figure~14 in][]{khoperskov22c}. This result hints at the importance of the individual galaxy assembly history in setting the timing of disk spin-up. 

In this paper, we aim to assess whether the discrepancy between the data and simulations points to some generic deficiency in modern galaxy formation models or is merely a result of statistical variation (i.e., the MW is just unusual). More broadly, we investigate how and when MW progenitor galaxies form their disks. In a companion paper \citep{semenov23b}, we also explore physical mechanisms governing the formation of galactic disks.

To address these questions, we use the TNG50 cosmological simulation, part of the IllustrisTNG project. 
Some aspects of disk formation and evolution have already been investigated in TNG50. For example, \citet{pillepich19} showed that TNG50 produces a disk population that agrees with the data reasonably well, e.g., in terms of disk sizes as a function of galaxy mass and redshift. \citet{pillepich23} focused on a subsample of MW and M31-like analogs and demonstrated that they exhibit significant galaxy-to-galaxy variation in star formation histories and disk sizes, morphologies, and kinematics, with the MW and M31 falling within the predicted scatter. \citet{sotillo-ramos22} also investigated the survival of such MW and M31-like disks through mergers and showed that the majority of these galaxies undergo at least one major merger in their history, with $\sim 20\%$ of these galaxies having such a merger within the last 5 Gyr but the disk either survives or is able to reform.

Here, we focus on a subsample of MW-like TNG50 galaxies (selected differently from \citealt{sotillo-ramos22} and \citealt{pillepich23}) and investigate the signatures of disk formation in the chemistry and kinematics of star particles, motivated by the local ``galactic archeology'' discoveries (see the references above). 

The paper is organized as follows. Section~\ref{sec:methods} briefly reviews the TNG50 simulation and outlines our methods and analyses. In Section~\ref{sec:results}, we present our key results. In particular, we show that the disk formation time and metallicity span wide ranges (Section~\ref{sec:results:dist}) and identify a population of early-forming galactic disks akin to our own MW (Section~\ref{sec:results:early-late}). In Section~\ref{sec:results:age-metallicity}, we present the implications of the scatter and shape of the age--metallicity relation for the correspondence between the disk spin-up signature in stellar age and [Fe/H]. Section~\ref{sec:results:assembly} shows that the scatter in the disk formation time is mostly associated with the scatter in halo mass assembly history. In Section~\ref{sec:results:protogalaxy}, we quantify the role of mergers during the protogalaxy stage. Finally, Section~\ref{sec:results:diskiness} shows that the disk properties at redshift $z=0$ are correlated with their formation timing, with earlier-forming disks being more disk-dominant. Section~\ref{sec:discussion} discusses the implications of our results and Section~\ref{sec:summary} summarizes our conclusions. In Appendix~\ref{app:metallicity}, we describe our method for metallicity calibration in TNG50 and investigate the sensitivity of our results to this choice.

\section{Methods}
\label{sec:methods}

\subsection{TNG50 Overview}

For our analysis, we use TNG50, the highest-resolution run of the IllustrisTNG cosmological simulation suite \citep{springel18,pillepich18b,nelson18,marinacci18,naiman18,nelson19b}. Here we outline the key features of this simulation; for more detail, refer to \citet{pillepich19},  \citet{nelson19}, and the TNG method papers \citep{weinberger17,pillepich18}.

The simulation was carried out using the quasi-Lagrangian $N$-body and magnetohydrodynamic (MHD) code Arepo \citep{arepo}. The code employs an unstructured moving mesh based on Voronoi tessellation of the simulation domain with a second-order-accurate unsplit Godunov-type method for hydrodynamic flux computation, an eight-wave Powell cleaning scheme to handle ideal MHD  \citep{pakmor11,pakmor13}, and a tree-particle-mesh method to solve the Poisson equation for the gravitational potential. 

The initial conditions for TNG50 are a representative $\sim 50^3$ comoving Mpc$^3$ volume of the Universe, assuming Planck XIII (\citeyear{planck-xiii}) cosmology: matter density $\Omega_{\rm m} = 0.3089$, baryon density $\Omega_{\rm b} = 0.0486$, cosmological constant $\Omega_\Lambda = 0.6911$, Hubble constant $H_0 = 67.74\;{\rm km\;s^{-1}\;Mpc^{-1}}$, and perturbation power spectrum with the normalization of $\sigma_8 = 0.8159$ and slope of $n_{\rm s} = 0.9667$. The box is selected out of 60 realizations, to produce the closest to average cumulative dark matter halo mass function for halos more massive than $10^{10} \Msun$ and is evolved from $z \sim 127$ to $z = 0$.

The mass resolution for dark matter in TNG50 is $4.5 \times 10^5\Msun$, while the target baryonic mass resolution is $8.5 \times 10^4\Msun$, with the actual mass of gas cells and stellar particles being maintained within a factor of two of this value. The resulting median cell size in MW progenitor galaxies at $z=1$ is $\sim 80\text{--}100\pc$ \citep[see Figure~1 in][]{pillepich19}. The gravitational softening length for dark matter, stellar, and wind particles is 575 comoving pc at $z>1$ and then fixed at 288 physical pc at $z<1$, while for gas, the softening length varies with the cell size as $\epsilon_{\rm gas} = 2.5\;r_{\rm cell}$, where $r_{\rm cell}$ is the radius of a sphere with the same volume as the Voronoi cell.

TNG50 models the key baryonic processes, including gas heating and cooling, star formation, supernova-driven galactic outflows, and active galactic nucleus (AGN) feedback. The simulation follows primordial, Compton, and metal-line cooling, with an approximate model for self-shielding corrections in the dense ISM, while the heating is provided by the spatially uniform \citet{faucher-giguere09} UV background and local AGN sources using the original Illustris model \citep{vogelsberger13}. To model the effects of local (i.e., unresolved) stellar feedback on the ISM, at densities $n > 0.1 \cc$, the gas pressure is replaced by the effective equation of state, following the \citet{sh03} model with the modifications described in \citet{vogelsberger13} and \citet{nelson19}. This subgrid model is also used to calculate the local star formation rate and  stochastically convert gas cells into star particles. Star formation-driven galactic winds are modeled via a kinetic wind approach, whereby hydrodynamically decoupled wind particles are stochastically ejected from the ISM with a prescribed mass loading at injection and redshift-dependent kick velocity and are recoupled with gas when they reach the background density of $n = 2.5 \times 10^{-3} \cc$, typically within few kiloparsecs from the star-forming ISM \citep{pillepich18,nelson19}. AGN feedback is modeled in two modes, based on the local Eddington-limited Bondi gas accretion rate: continuous injection of thermal energy at high accretion rates and stochastic kinetic kicks at low accretion rates \citep{weinberger17}.

TNG50 tracks 9 chemical elements (H, He, C, N, O, Ne, Mg, Si, and Fe) plus the total amount of all other, untracked elements. The enrichment model accounts for Type II and Type Ia supernovae and asymptotic giant branch (AGB) stars channels with the tabulated yields described in Section~2.3.4 \citet{pillepich18}, assuming a \citet{chabrier03} IMF \citep[for more details, see][]{naiman18,pillepich18}. To alleviate the uncertainties in the assumed yields, for our analysis, we recalibrate metallicity values, as described in Section~\ref{sec:methods:analysis} and Appendix~\ref{app:metallicity}.

\subsection{Milky Way-like Galaxy Sample}
\label{sec:methods:sample}

\begin{figure}
\centering
\includegraphics[width=\columnwidth]{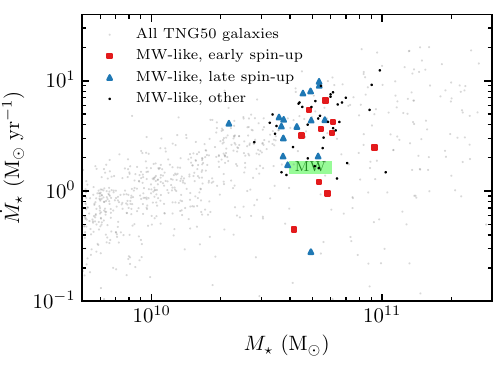}
\caption{\label{fig:Mstar-SFR} Sample of MW-like TNG50 galaxies in the stellar mass--SFR plane. The bright colors show 61 galaxies that pass the selection criteria described in Section~\ref{sec:methods:sample}, while the faint gray points show all TNG50 galaxies. The MW-like sample is further subdivided based on the timing of the disk formation (see Section~\ref{sec:results:early-late} for detail) into early-spin-up (red squares), late-spin-up (blue triangles), and the rest (black dots). The Milky Way values with the associated uncertainties are indicated with a green square \citep{mw-review}. Note that the simulation points show the total values for the FoF groups; for the majority of galaxies from our sample, the $M_{\rm \star}$ of FoF are dominated by that of the central galaxy, while the SFRs are either dominated or within a factor of $\sim 2$ of that of the central galaxy.}
\end{figure}

\begin{figure*}
\centering
\includegraphics[width=0.85\textwidth]{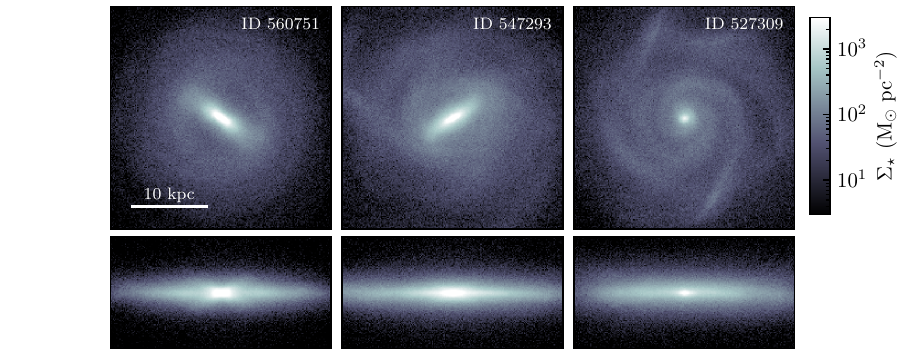}\\
\includegraphics[width=0.85\textwidth]{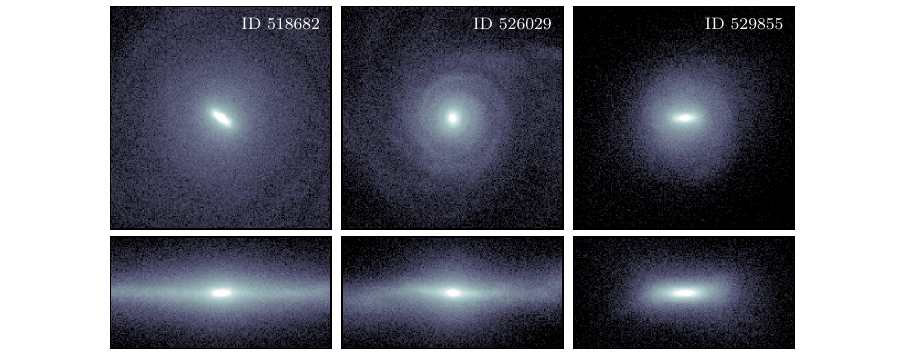}\\
\caption{\label{fig:maps} Face-on and edge-on views of stellar disks of MW-like TNG50 galaxies selected using the criteria from Section~\ref{sec:methods:sample}. The top and bottom rows show examples of early- and late-spin-up galaxies, respectively, with the former being close MW analogs in the chemokinematic signatures of their stars (see Section~\ref{sec:results:diskiness} for details). The MW-like galaxies in TNG50 exhibit a wide variation of morphology in terms of disk size and the presence and strength of bulges, bars, and spiral arms.}
\end{figure*}

To select a sample of MW-like galaxies, we apply the following criteria to the population of TNG50 central (i.e., not satellite) galaxies at redshift $z=0$:

\begin{itemize}
    \item \emph{Milky Way-mass dark matter halo.} $M_{\rm 200c} = 0.8\text{--}1.4 \times 10^{12}\Msun$, where $M_{\rm 200c}$ is the mass of the friends-of-friends (FoF) group defined based on the spherical overdensity of 200 with respect to the critical density of the Universe. 
    This range of $M_{\rm 200c}$ is motivated by the average estimate from a compilation of halo stellar kinematic studies from \citet{mw-review} and is also consistent with more recent estimates \citep[e.g.,][]{zaritsky20,deason21,vasiliev21,correamagnus22}.

    \item \emph{Star-forming.} To exclude quiescent galaxies, we require the FoF group hosting the galaxy to have a total instantaneous SFR of at least $\SFR > 0.2\Msunyr$. For the majority of galaxies from our resulting sample, this SFR is within a factor of $\sim 2$ from the SFR of the galaxy itself.

    \item \emph{Disky.} We require that the young ($<100$ Myr old) stellar disk is rotationally supported with $\vsyoung > 6$, where we use median $\vrot$ computed for young ($<100$ Myr old) stars within twice the radius containing half of the total SFR. The 1D velocity dispersion $\sigma$ is computed as half of the 16--84 interpercentile range of $\vrot$. In addition, we also excluded four galaxies with significantly perturbed disks at $z=0$.
\end{itemize}

In total, 61 TNG50 galaxies pass the above criteria, which we will refer to as ``MW-like galaxies'' henceforth. Figure~\ref{fig:Mstar-SFR} shows the locations of these galaxies in the stellar mass--SFR plane, with the MW estimates shown with a green rectangle. The stellar mass contained in the selected halos lies within a factor of 2 of the MW estimates. Most of the MW-like galaxies lie on the main sequence, but there is also a substantial tail of low-SFR green-valley galaxies. Note that the simulation points show the total values for the FoF groups and therefore should not be compared with the MW values literally. However, we checked that for the majority of galaxies from our sample, the $M_{\rm \star}$ of the groups are dominated by that of the central galaxy, while the SFRs are either dominated or within a factor of $\sim 2$ of that of the central galaxy.
\newtext{In addition, although our sample is selected based on the presence of a relatively strong young stellar disk, the rotational supports of the total stellar disks are also consistent with the typical observed values, $\vsall \sim 2\text{--}5$ \citep[][see Figure~\ref{fig:diskiness} below]{kregel05}. }

In Section~\ref{sec:results:early-late} below, we further subdivide our galaxy sample into ``early-'' and ``late-spin-up'' galaxies based on the stellar metallicities of the spin-up feature introduced by \citetalias{bk22}, which marks the timing of disk formation. For reference, these subsamples are also indicated in Figure~\ref{fig:Mstar-SFR}. Interestingly, the early-spin-up galaxies---which are close MW analogs in terms of the chemokinematic distributions of their stars---tend to lie below the main sequence, also in agreement with the MW being a green-valley galaxy. 

Although, by design, all galaxies from our MW-like sample are disky, they show a significant morphological diversity, as illustrated in Figure~\ref{fig:maps}. The top and bottom rows show examples of early- and late-spin-up galaxies, respectively. Early-spin-up galaxies tend to be more disk-dominated, while late-spin-up ones can have more pronounced bulges (see Section~\ref{sec:results:diskiness} for more quantitative analysis). The morphologies of these galaxies also vary by the presence and strength of the galactic bars and spiral arms.

\subsection{Analysis}
\label{sec:methods:analysis}

In this subsection, we briefly outline the key choices and definitions used in our analysis.

\emph{Spatial selection of stellar particles.} In what follows, we will use the chemistry and kinematics of stellar particles from TNG50 to investigate the imprint of galactic disk formation in the chemokinematic data of MW stars, using the results of \citetalias{bk22} as an observational reference. This observational sample of nearby stars extends $\sim 3$ kpc away from the Sun, and therefore, for each MW-like TNG50 galaxy, we select in situ stars within a range of galactocentric $R = 5\text{--}11$ kpc and a distance from the disk midplane of $|z| < 3$ kpc.

\emph{Selection of in situ stars.} While stellar metallicities serve to identify generally old stellar populations, stars brought by galaxy mergers introduce significant scatter into this identification, as they are typically formed in lower-metallicity environments. To include only in situ stars in our analysis, we select stars that were formed within 30 comoving kiloparsecs from the main progenitor.

\begin{figure}
\centering
\includegraphics[width=\columnwidth]{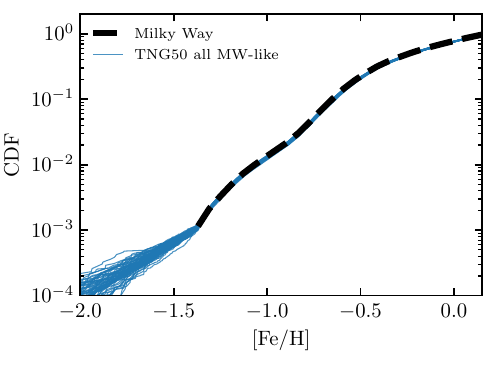}\\
\includegraphics[width=\columnwidth]{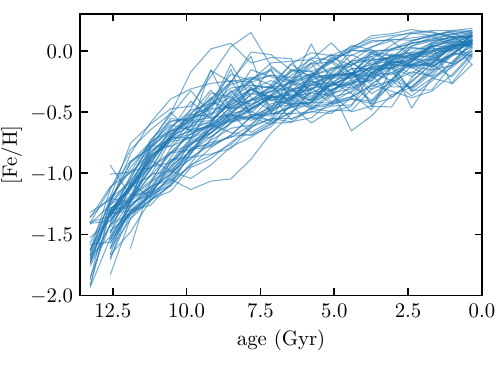}
\caption{\label{fig:CDF-FeH-age} For our analysis, we recalibrate the values of [Fe/H] of in situ star particles by abundance matching the CDF from the observational sample of \citetalias{bk22} while preserving the original order of [Fe/H] predicted in TNG50. The top panel compares the resulting CDFs of [Fe/H] for all of our TNG50 MW-like galaxies (thin blue lines) with the observed CDF \citepalias[dashed black line;][]{bk22}. The bottom panel shows the median [Fe/H] (after recalibration) as a function of star particle ages.}
\end{figure}

\emph{Metallicity calibration.} Assumptions about metal yields and stellar evolution models constitute one of the key uncertainties for predicting metallicity distributions in galaxy simulations. To mitigate this problem, we recalibrate the metallicities of the stellar particles selected above to exactly match the cumulative distribution function (CDF) of [Fe/H] reported in \citetalias{bk22} using a method analogous to abundance matching. This calibration preserves the original order of the stellar metallicities produced in TNG50 and only rescales their values. The resulting distribution of [Fe/H] and the age--metallicity relation are shown in Figure~\ref{fig:CDF-FeH-age}. In Appendix~\ref{app:metallicity}, we also compare our results with other choices for metallicity calibration, specifically the raw [Fe/H] predicted in TNG50 and the constant shift used in \citetalias{bk22}. 
\newtext{In particular, we show that raw [Fe/H] exhibit shallower CDFs than observed in the MW, implying a larger relative abundance of low-metallicity stars. Our calibration, therefore, effectively leads to earlier enrichment compared to the raw TNG50 results, which pushes the spin-up feature to $\sim 0.1\text{--}0.2$ dex higher [Fe/H] and makes this transition sharper.}

\emph{Galaxy orientation.} The rotational velocity and the angular momentum vector are defined with respect to the frame in which the $z$-axis is directed along the angular momentum vector of the existing stellar population.

\emph{Orbital circularity.} In Figure~\ref{fig:diskiness} below, we use the distribution of the orbital circularities of star particles, $j_z/j_{\rm c}$, as a measure of galaxy diskiness. $j_{\rm c}$ is defined for each star particle as the specific angular momentum of the circular orbit having the same energy as the particle energy, $E = \phi + |v|^2/2$, where $\phi$ is the gravitational potential and $|v|$ is the magnitude of particle velocity with respect to the galaxy. To this end, we use a spherically averaged $\phi$ and invert the relation between orbital energy and radius, $E_{\rm c}(R_{\rm c}) = \phi(R_{\rm c}) + v_{\rm c}(R_{\rm c})^2/2$, with $v_{\rm c}^2 = R\;d\phi/dR$, to compute $j_{\rm c} \equiv R_{\rm c} \; v_{\rm c}(R_{\rm c})$ at $R_{\rm c}(E_{\rm c} = E)$. By definition, $j_z/j_{\rm c}$ can take values from 1 (corotation with the galaxy on a circular orbit) to -1 (counter-rotation), with 0 corresponding to the motion along radial or polar orbits.

\section{Galactic Disk Formation in TNG50}
\label{sec:results}

\subsection{The MW is Unusual, but Consistent with the Population of MW-mass Galaxies}
\label{sec:results:dist}

\begin{figure}
\centering
\includegraphics[width=\columnwidth]{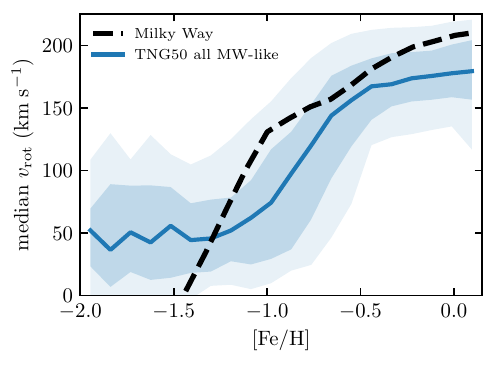}\\
\includegraphics[width=\columnwidth]{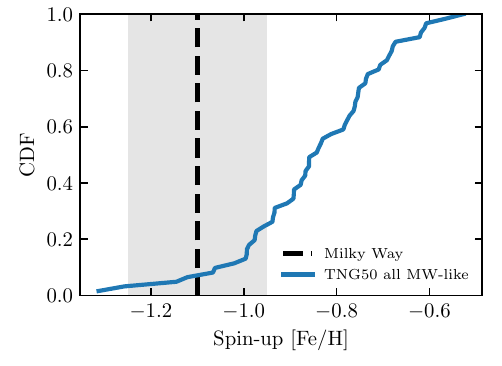}
\caption{\label{fig:FeH-vrot-median} Comparison of the disk formation imprint in stellar chemistry and kinematics between TNG50 and MW measurements. {\bf Top:} median rotational velocity of stellar particles as a function of metallicity for our sample of 61 MW-like TNG50 galaxies. The relation is computed for each galaxy individually, and the thick blue line shows the median relation, while the shaded regions show 16--84 and 2.5--97.5 interpercentile ranging, roughly corresponding to 1 and $2\sigma$ scatter. The thick dashed line shows the relation measured in the MW by \citetalias{bk22}. {\bf Bottom:} CDF of stellar metallicities at which the spin-up feature occurs. This metallicity is defined as the value of [Fe/H] at which median $\vrot$ equals half that at solar metallicity. The vertical dashed line indicates the spin-up [Fe/H] $\approx -1.1$ measured in the same way for the MW, while the shaded band shows the extent of [Fe/H] over which the spin-up occurs, $\sim$0.3 dex. TNG50 shows a broad variation of [Fe/H] at which the galactic disk spins up, and the MW is consistent with this distribution, being on the low-metallicity side $\sim 2\sigma$ away from the median.}
\end{figure}

We start by comparing the relation between stellar metallicity and $\vrot$ for all stars in the population of MW-like galaxies from TNG50 with the relation found in the MW by \citetalias{bk22}. To approximate the analysis of \citetalias{bk22}, we calculate the $\vrot$ of each in situ star particle within $R = 5\text{--}11$ kpc and $|z| < 3$ kpc in the reference frame aligned with the angular momentum of the stellar disk and find the running median of $\vrot$ as a function of the recalibrated [Fe/H] for each of the 61 MW analogs from our TNG50 sample (see Sections~\ref{sec:methods:sample}--\ref{sec:methods:analysis} and Appendix~\ref{app:metallicity} for details). The top panel of Figure~\ref{fig:FeH-vrot-median} shows the median relation stacked across these simulated MW analogs together with the interpercentile ranges that roughly correspond to $1\sigma$ and $2\sigma$ galaxy-to-galaxy variation around the median.

The shape of the simulated relation is qualitatively similar to the observed one, with a low net rotation at low metallicities and a sharp spin-up of $\vrot$ at a higher metallicity. The range of metallicities at which the spin-up occurs is quite wide, $\sim$0.6 dex, and the median value is shifted to a $\sim$0.3 dex higher metallicity than observed, analogous to the late spin-up in the FIRE and Auriga simulations reported by \citetalias{bk22}. The observed relation, however, falls within the $\sim2\sigma$ scatter of the predicted distribution, indicating that \emph{although the MW is unusual, it is consistent with the overall population of MW-mass disk galaxies}.

This consistency is further quantified in the bottom panel of Figure~\ref{fig:FeH-vrot-median}, which shows the cumulative distribution of the spin-up metallicity, defined as the value of [Fe/H] at which the median rotational velocity of stars is approximately half of that at solar metallicity. Again, roughly 10\% of MW-like TNG50 galaxies show the spin-up feature at lower metallicity than observed in the MW ([Fe/H] $\sim -1.1$; shown with the vertical dashed line). As we will show in the next subsection, such galaxies exhibit a relation between $\vrot$ and [Fe/H] that closely resembles MW observations.

\subsection{Early- and Late-spin-up Galactic Disks}
\label{sec:results:early-late}

\begin{figure*}
\centering
\includegraphics[width=0.8\textwidth]{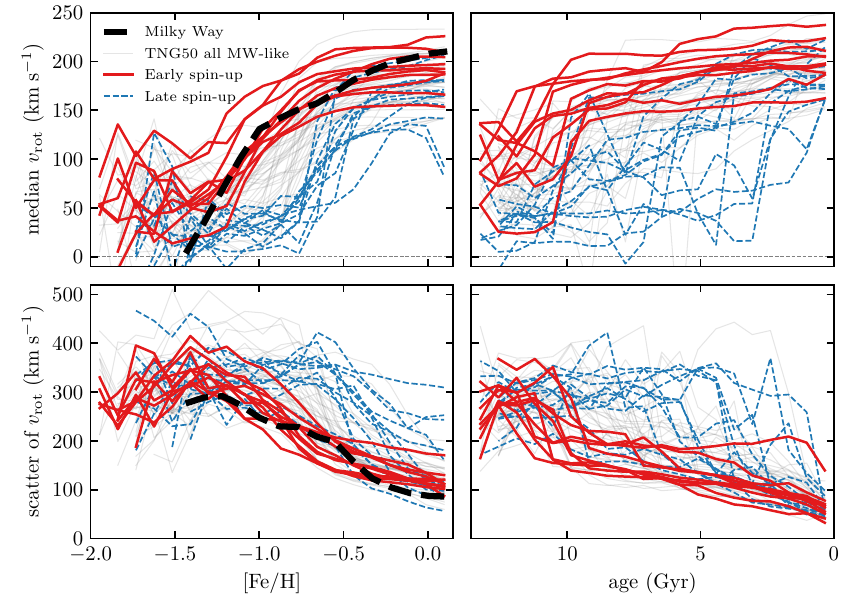}
\caption{\label{fig:FeH-vrot} Relations between stellar metallicities, ages, and kinematics for individual MW-like TNG50 galaxies. The top and bottom rows show the median rotational velocities of star particles and their scatter, respectively. The scatter is measured as the 5--95 interpercentile range for direct comparison with \citetalias{bk22} observations (black dashed lines). The left and right columns show these quantities as functions of metallicities and stellar ages, respectively. The colored lines highlight subsamples of simulated galaxies that show a sharp spin-up of $\vrot$ either at the lowest (early-spin-up; solid red lines) or highest [Fe/H] (late-spin-up; dashed blue lines). The two samples are clearly separated in all panels, with the early-spin-up galaxies having relations similar to that observed in the MW.}
\end{figure*}

The above results show that the metallicity at which the spin-up occurs in simulated galaxies covers a wide range of values, $-1.2 \lesssim \text{[Fe/H]} \lesssim -0.5$, with the value observed in the MW falling within the lower end of this range. A natural question arises: what creates this wide variation? Why do some galactic disks (like our MW) spin up early, while others spin up late?

To gain further insights into this question, Figure~\ref{fig:FeH-vrot} shows the relations for individual MW-like TNG50 galaxies, with the highlighted galaxies showing the spin-up feature at low metallicities, closely resembling the MW in the [Fe/H]--$\vrot$ plane (red lines in the plot), and the ones for which the spin-up occurs at the highest metallicities (dashed blue lines). In what follows, we dub these two samples as \emph{early-spin-up} and \emph{late-spin-up} galaxies, respectively. The early-spin-up galaxies are close MW analogs in terms of their chemokinematic signatures.
These two subsamples were selected by eye as the galaxies lying at the opposite ends of the distribution and, in the case of early-spin-up galaxies, showing a sharp spin-up feature qualitatively similar to the MW; using a formal criterion to identify these two subsamples would not qualitatively affect any of the conclusions.

The top panels show the median rotational velocity as a function of stellar metallicities (left) and ages (right). While the two samples are defined solely based on their location in the [Fe/H]--$\vrot$ plane (top left panel), the figure shows a clear separation between these samples also in stellar ages, although some overlap appears (top right panel). The early-spin-up galaxies form early ($\gtrsim10$ Gyr ago) and rapidly (over $\sim1\text{--}2$ Gyr). The late-spin-up galaxies, on the other hand, show a broader diversity, both in terms of the sharpness and the stellar age at which the spin-up occurs. As expected, the spin-up feature generally occurs at younger ages than in the early-spin-up sample. However, a couple of late-spin-up galaxies formed a disk $\sim 10$ Gyr ago, indicating that these galaxies have already been highly enriched by that time, causing their spin-up feature to occur at high metallicity, [Fe/H] $\gtrsim -0.7$.

Interestingly, a similar dichotomy between early- and late-spin-up galaxies is also clear in the scatter of $\vrot$ shown in the bottom panels. Early-spin-up galaxies exhibit elevated scatter of $\vrot$ ($\sim 250\text{--}350\kms$; \newtext{calculated as a 5--95 interpercentile range for direct comparison with the \citetalias{bk22} results}) for low-metallicity ([Fe/H] $<-1$) stars older than $\sim$10 Gyr, corresponding to the pre-spin-up epoch. For higher-metallicity and younger stars, the scatter drops to $\sim 100\text{--}200\kms$. The overall magnitude of the scatter in $\vrot$ and its dependence on metallicity is in a broad agreement with the observations of \citetalias{bk22} shown with the black dashed line. The late-spin-up galaxies, in contrast, generally exhibit elevated scatter down to ages $\lesssim 5$ Gyr and solar metallicity, although there are several exceptions where late-spin-up galaxies show a scatter in $\vrot$ similar to the early-spin-up sample.
Overall, the $\vrot$ scatter shows the reverse behavior compared to the median $\vrot$: it is high early on and decreases shortly after the disk spin-up, at $\sim 0.3$ dex higher [Fe/H] than the spin-up value.

\begin{figure}
\centering
\includegraphics[width=\columnwidth]{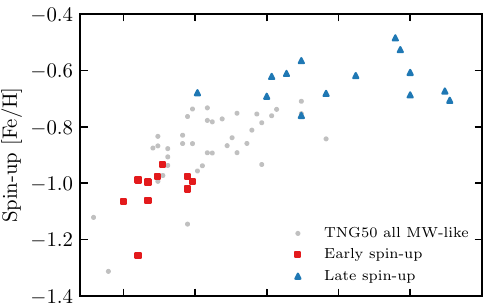}\\
\includegraphics[width=\columnwidth]{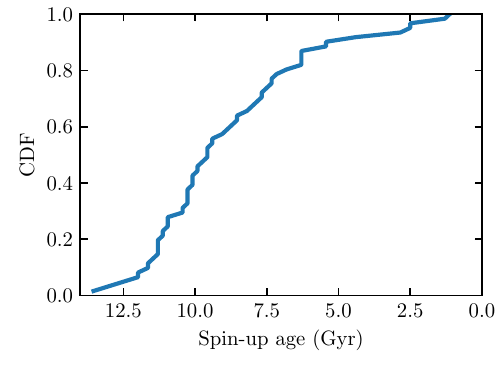}
\caption{\label{fig:spinup-age} {\bf Top:} the relation between stellar ages and [Fe/H] at which the spin-up feature occurs in the top panels of Figure~\ref{fig:FeH-vrot}. The red squares and blue triangles indicate early- and late-spin-up galaxies, respectively. This relation is a subsample of the overall age--metallicity relation, and therefore it has qualitatively the same shape as the latter, with the early rapid rise and late slow increase (cf. Figure~\ref{fig:CDF-FeH-age}). This shape explains why late-spin-up galaxies exhibit significant variation of spin-up ages and sharper transitions in the [Fe/H]--median $\vrot$ space (see the text). {\bf Bottom:} cumulative distribution of spin-up ages. Roughly $40\%$ of galaxies form as early as the early-spin-up disks ($>10$ Gyr ago) and yet only 1/4 of these galaxies exhibit the spin-up feature at [Fe/H] $\lesssim -1$ as the MW (see Figure~\ref{fig:FeH-vrot-median}). This is a direct consequence of the rapid rise and significant scatter of [Fe/H] at these early times.}
\end{figure}

\subsection{Implications of the Shape and Scatter in the Age--Metallicity Relation}
\label{sec:results:age-metallicity}

The significantly wider variation of the disk spin-up time for the late-spin-up galaxies compared to the early ones is easy to understand from the relation between $\tspinup$ and the corresponding [Fe/H]. This relation is shown in the top panel of Figure~\ref{fig:spinup-age}, where we defined the spin-up age analogously to the spin-up [Fe/H]: the stellar age at which median $\vrot$ is approximately half of its final value. This relation has the same shape as the overall age--metallicity relation (recall Figure~\ref{fig:CDF-FeH-age}): [Fe/H] rapidly rises at ages $>10$ Gyr and flattens after that. The lowest metallicities of the early-spin-up disks (red squares) are reached during the early rapid enrichment,\footnote{Note that there are several galaxies with low spin-up metallicity ([Fe/H] $<-1$) that were not marked as early-spin-up ones; this is because these galaxies show a very shallow rise of median $\vrot$ vs. [Fe/H] instead of a sharp spin-up, as in the MW.} while the metallicities of the late-spin-up disks (blue triangles) are all in the slow enrichment regime. As a result, $\sim$ 0.3 dex change in the [Fe/H] of the early-spin-up galaxies translates into a narrow variation of ages, $\sim12.5\text{--}10$ Gyr, while the same variation in the [Fe/H] of late-spin-up disks corresponds to a significant range of ages, $\sim10\text{--}1$ Gyr.

Another consequence of the age--metallicity relation is the difference in the sharpness of the spin-up feature: during the spin-up, stellar [Fe/H] in early-spin-up galaxies changes more gradually compared to the sharp rise in the late-spin-up disks (see the top left panel of Figure~\ref{fig:FeH-vrot}). Again, this is because the physical timescale of disk formation translates into a significantly larger variation of [Fe/H] during the early, rapid enrichment phase than during the late, slow enrichment regime.

Finally, the early rapid rise of [Fe/H] and its scatter also lead to a significant fraction of galaxies that spin up early ($\gtrsim 10$ Gyr ago), but are already highly enriched, [Fe/H] $>-1$. This is clear from the comparison of the CDF of spin-up ages in the bottom panel of Figure~\ref{fig:spinup-age} with the CDF of the spin-up [Fe/H] in Figure~\ref{fig:FeH-vrot-median}. While only $\sim 10\%$ of MW-mass disks spin up at metallicities [Fe/H] $\lesssim -1$ (like the MW), $\sim 40\%$ of galaxies spin up at the same ages as these MW analogs, i.e., $\gtrsim 10$ Gyr ago. In other words, only 1/4 of early-forming disks exhibit the spin-up feature at as low metallicities as observed in the MW, further making the MW unusual.

\subsection{Disk Formation Is Correlated with Mass Assembly}
\label{sec:results:assembly}

\begin{figure}
\centering
\includegraphics[width=\columnwidth]{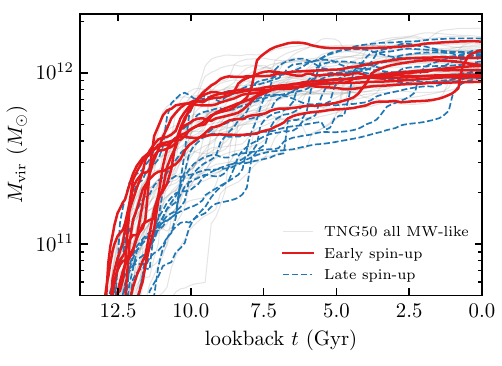}
\caption{\label{fig:Mvir} Mass assembly histories of MW-like TNG50 galaxies. The dark matter halo virial mass, $\Mvir$, is computed using the redshift-dependent spherical overdensity from \citet{bryan-norman98}. Halos hosting early-spin-up galaxies assemble most of their mass early and rapidly, while hosts of late-spin-up galaxies generally assemble over more extended periods and go through significant mergers (manifested as vertical jumps of $\Mvir$). Thus, the shape of the evolution of $\Mvir$ is qualitatively analogous to the tracks of the median rotational velocity as a function of stellar ages in the upper right panel of Figure~\ref{fig:FeH-vrot}. }
\end{figure}

As we will show in this subsection, the above trends and differences between early- and late-spin-up galaxies can be explained by the differences in their mass assembly histories, in particular, the occurrence of recent major mergers.

Figure~\ref{fig:Mvir} shows the evolution of the host dark matter halo virial mass (defined using the \citealt{bryan-norman98} fit for the spherical overdensity) for our sample of MW-like TNG50 galaxies. On average, halos hosting early-spin-up galaxies (red lines) quickly assemble a significant fraction of their mass early on, during the first few Gyr of evolution, and then grow steadily over the last $\sim 10$ Gyr. In contrast, late-spin-up galaxy hosts (dashed blue lines) assemble their mass over a longer timescale and, in detail, show a variety of mass assembly tracks: some grow gradually over extended periods, some show a single or a series of sudden jumps of $\Mvir$, corresponding to major mergers, and a few form early, analogous to the early-spin-up galaxies.
This behavior is reminiscent of the relation between the median $\vrot$ and stellar ages discussed in the previous section (the upper right panel of Figure~\ref{fig:FeH-vrot}), where early-spin-up stellar disks form early and rapidly, while the late-spin-up disks generally form later, with a significant variation in timing and duration.

\begin{figure*}
\centering
{\centering \large \bf Early spin-up galaxy}
\includegraphics[width=0.8\textwidth]{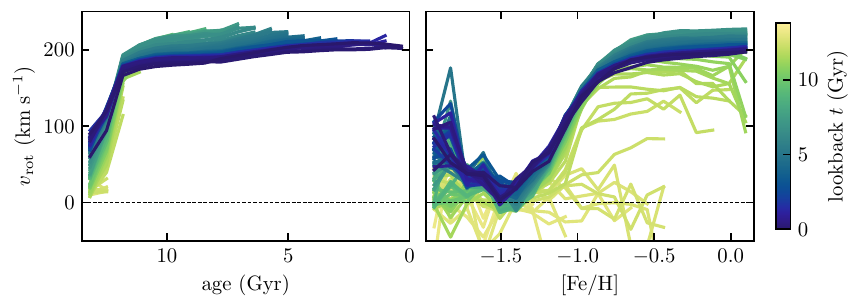}\\
\vspace{0.5em}
\includegraphics[width=\textwidth]{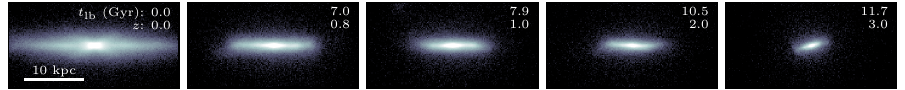}\\
\vspace{3em}
{\centering \large \bf Late spin-up galaxy}
\includegraphics[width=0.8\textwidth]{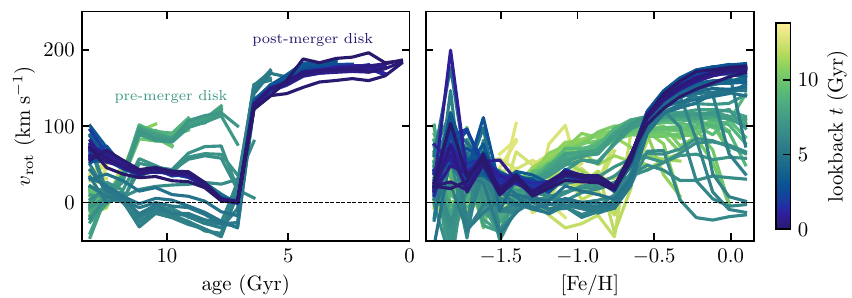}\\
\vspace{0.5em}
\includegraphics[width=\textwidth]{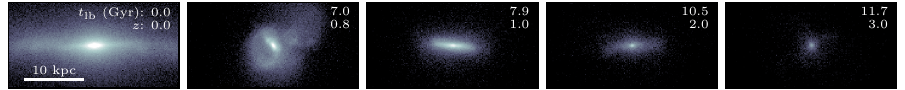}
\caption{\label{fig:early-late-examples} Examples of the evolution of median $\vrot$ vs. stellar age (left) or metallicity (right) for early- (top) and late-spin-up galaxies (bottom). The maps at the bottom of each set of panels show the edge-on view of the stellar disk (defined based on the net angular momentum direction) at different redshifts (indicated in the panels together with the lookback time, $t_{\rm lb}$); the color range is the same as in Figure~\ref{fig:maps}. The disk of an early-spin-up galaxy is in place early on, $\sim$12 Gyr ago (i.e., by $z \approx 3.5$), and steadily evolves until the present day. The late-spin-up galaxy also forms a disk early on, around the same time as the early-spin-up one (labeled as the ``pre-merger disk'' in the plot). However, a major merger $\sim$7 Gyr ago ($z \approx 0.8$; dark matter mass ratio of $\sim$ 1:2) destroys that disk, and another disk forms shortly after. As a result, this merger resets the spin-up feature to the age and metallicity of stars corresponding to the timing of that merger.}
\end{figure*}
 
To elucidate the connection between the mass assembly history and disk spin-up, Figure~\ref{fig:early-late-examples} shows two representative examples of the evolution of median $\vrot$ as a function of stellar age and metallicity for an early- (top) and a late-spin-up galaxy (bottom). To construct these plots, we use the ages and metallicities of the star particles selected in the final snapshot and recompute their $\vrot$ in each previous snapshot (shown with different colors) with respect to the orientation of the stellar disk in that snapshot.

For early-spin-up galaxies (the top row in Figure~\ref{fig:early-late-examples}), the relation between median $\vrot$ and stellar age and metallicity is built up early on and then evolves gradually, without sudden changes. The spin-up feature is in place after a few Gyr of evolution and does not change significantly over the last 12 Gyr. This is especially clear in the left panel, showing stellar ages: later snapshots simply add progressively younger stars, while the evolution of stellar metallicities is more complicated, as newly formed stars always have a range of metallicities.

The evolution of late-spin-up galaxies is qualitatively different, in particular, due to the presence of significant mergers. The bottom panels of Figure~\ref{fig:early-late-examples} show a late-spin-up galaxy that experienced a major merger $\sim 7$ Gyr ago with a $\sim$ 1:2 dark matter mass ratio at the infall. At lookback $t \gtrsim 7$ Gyr, this galaxy had a stellar disk with $\vrot \sim 100\kms$ (labeled the ``pre-merger disk'' in the plot) that was formed within the first few Gyr of evolution and had the spin-up feature at ages $\sim 11$ Gyr and [Fe/H] $\sim -1.2$, just like the early-spin-up galaxy shown in the top panels. The major merger at $t \sim 7$ Gyr destroyed this early disk, and shortly after, the new disk formed from the gas left after the merger. In this galaxy, the spin-up feature marks the age and metallicity corresponding to the timing of that merger.

This example shows that the main effect of a destructive major merger is ``resetting'' the spin-up feature to the time and metallicity of the merger. Correspondingly, the resulting distributions of $\vrot$ vs. stellar age or metallicity will show significant variation, depending on the timing, mass ratio, relative orbits, and number of mergers.
In addition, the range of ages at which the spin-up feature occurs in late-spin-up galaxies is further increased due to the shallow age--metallicity relation (Section~\ref{sec:results:age-metallicity}).
We find that the majority of galaxies from our sample experience a merger event with a dark matter mass ratio of at least 1:10 within 1 Gyr from the moment at which the spin-up feature occurs in the stellar ages. 

It is worth noting, however, that the spin-up feature does not generally correspond to the last merger event, because the disk can survive through such mergers \citep[see also][]{sotillo-ramos22}. Indeed, several of the early-spin-up galaxies in Figure~\ref{fig:Mvir} exhibit jumps in $\Mvir$ at late times, indicating significant mergers (see such jumps at lookback $t \sim 7.5$, $4$, and $1$ Gyr).
Their stellar disks, however, survive through these mergers. A similar nondestructive merger occurred in the MW, albeit at earlier times: Gaia Sausage/Enceladus (GSE) only ``splashed'' some of the stars into the halo without destroying the pre-existing disk \citepalias{bk22}.

Interestingly, some of the late-spin-up galaxies have mass assembly histories (and age--$\vrot$ relations; see the top right panel of Figure~\ref{fig:FeH-vrot}) similar to the early-spin-up galaxies, but nevertheless they are classified as late-spin-up, because this feature occurs at high metallicity, [Fe/H] $\gtrsim -0.8$. Such galaxies were in the upper envelope of the age--metallicity relation when their disk formed (see Figure~\ref{fig:spinup-age}; in particular, one late-spin-up galaxy at $\sim 10$ Gyr and several more scattered around $\sim 7.5$ Gyr). Such galaxies are extreme examples of the early-forming disks, whose spin-up features in [Fe/H] vs. $\vrot$ space were upscattered to high [Fe/H] due to the scatter in the age--metallicity relation (Section~\ref{sec:results:age-metallicity}).

To summarize, early-spin-up galaxies, analogous to our own MW, rapidly assemble a significant fraction of their mass and form a disk early on, within the first few Gyr of evolution. To produce the spin-up feature at [Fe/H] $\lesssim -1$ like in the MW, the disk must form during the early, rapid enrichment phase (recall Figure~\ref{fig:spinup-age} and related text). After that, their mass grows steadily, without significant mergers that can disrupt the disk, and the disk therefore also grows steadily. 

While early-spin-up galaxies form alike, late-spin-up galaxies form in their different ways. \newtext{Significant} mergers can destroy pre-existing disks, resetting the spin-up feature to the timing and metallicity of that merger. The distributions of $\vrot$ as a function of stellar ages and metallicities will therefore exhibit significant galaxy-to-galaxy variation, due to the difference in the timing, mass ratio, and number of mergers. In addition, some of the late-spin-up galaxies can in fact be early-forming and have quiescent mass assembly histories, but their spin-up feature can be upscattered to high metallicity due to the scatter in the age--metallicity relation.

\subsection{Protogalaxy and Mergers during Early Assembly}
\label{sec:results:protogalaxy}

\begin{figure*}
\centering
{\centering \large Protogalaxy at $\tspinup-0.5$ Gyr}
\includegraphics[width=0.8\textwidth]{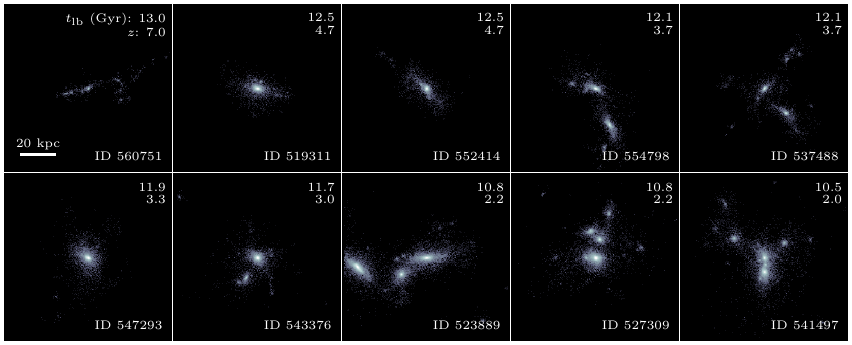}\\
\vspace{1em}
{\centering \large Same stars at $\tspinup+3$ Gyr}
\includegraphics[width=0.8\textwidth]{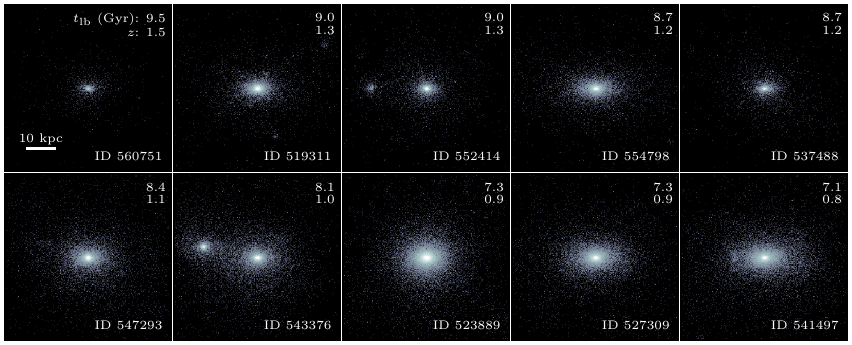}\\
\vspace{1em}
{\centering \large Stars formed between $\tspinup-0.5$ and $+3$ Gyr}
\includegraphics[width=0.8\textwidth]{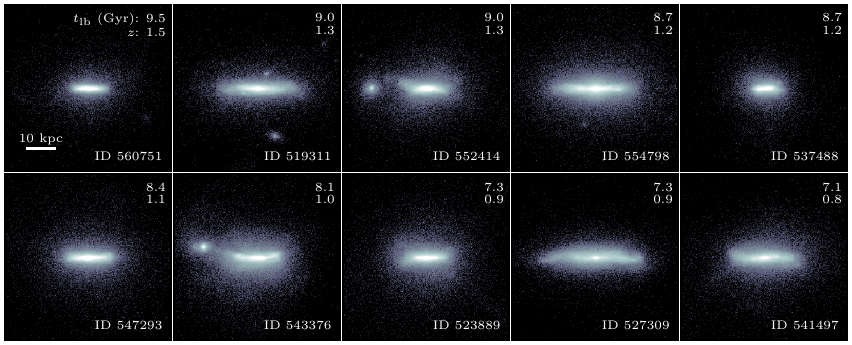}\\
\caption{\label{fig:maps-proto} Early-spin-up galaxies before and after disk formation. Galaxies are ordered by the disk formation time, with the lookback time and redshift indicated in the corner. All panels show edge-on views: the frame is oriented such that the total angular momentum of all stars bound to the main progenitor points upward. {\bf Top:} protogalaxy stage 0.5 Gyr prior to $\tspinup$, when disk formation sets in. {\bf Middle:} distribution of the same stars after the disk is established, 3 Gyr after $\tspinup$. {\bf Bottom:} stars that were formed between the snapshots shown in the top and middle sets, i.e., between $\tspinup - 0.5$ and $+ 3$ Gyr. The normalization of the color map is the same in the top two sets and is scaled down by a factor of 3 in the bottom set for presentation purposes. The early protogalaxy stage is associated with active mergers. The remnant of this stage becomes arranged in a bulge-like component around which the disk forms.}
\end{figure*}

\begin{figure}
\centering
\includegraphics[width=\columnwidth]{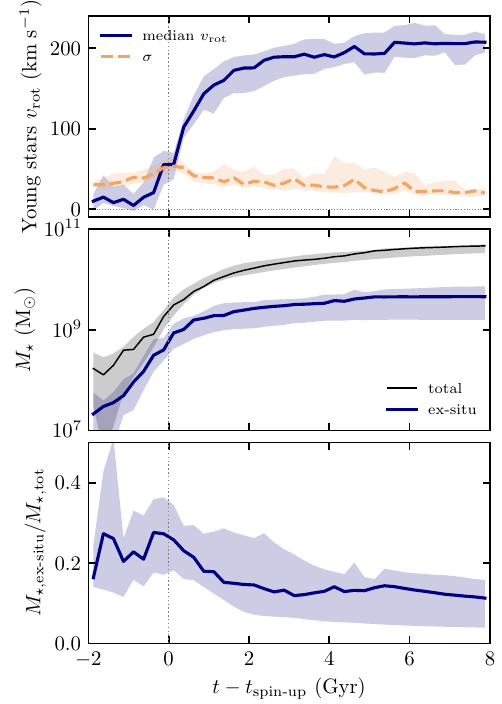}
\caption{\label{fig:f-exsitu} The growth of stellar mass and ex situ contribution via mergers before, during, and after disk formation. Evolution tracks are stacked among our early-spin-up galaxies, with the time axis shifted such that 0 corresponds to the time of disk spin-up, $\tspinup$, also shown with the vertical dotted line. The lines show median tracks and the shaded regions indicate galaxy-to-galaxy variation (16$^\text{th}$--84$^\text{th}$ interpercentile range). {\bf Top:} median $\vrot$ and its $1\sigma$ scatter of the young ($<100$ Myr) stellar disk. The formation of a young rotation-dominated disk coincides with $\tspinup$. {\bf Middle:} growth of the total stellar mass and ex situ mass brought in by mergers.  {\bf Bottom:} evolution of the instantaneous ex situ mass fraction. During the onset of disk formation, the ex situ contribution is substantial, $\sim20\%\text{--}30\%$, corresponding to the equivalent merger mass ratio between $\sim$1:4 and 1:2.}
\end{figure}

Similar to destructive mergers resetting the disk formation of late-spin-up galaxies, the disk formation of many early-spin-up galaxies is also marked by the end of the early assembly phase with ubiquitous significant mergers. Even though during this epoch one of the galaxies always has the largest mass, such frequent mergers make the notion of a single clear progenitor rather ambiguous and complicate its identification in the population of the most metal-poor stars observed in the Galaxy at the present day. To circumvent this possible ambiguity, \citet{conroy22} collectively termed all star formation occurring during this pre-disk epoch as the protogalaxy.

The top set of panels in Figure~\ref{fig:maps-proto} visualizes this protogalaxy stage for all of our early-spin-up galaxies. The panels show the stellar distribution 0.5 Gyr before the moment when the disk spins up ($\tspinup$, determined by the spin-up age defined in Section~\ref{sec:results:age-metallicity}), ordered by increasing $\tspinup$, with the lookback time and redshift indicated in the corner. As described above, most of the progenitors during this epoch are associated with active mergers. In some cases, the merging galaxies already have disks, e.g., the third panel in the second row.

The middle set of panels shows the distribution of the same stars, but 3 Gyr after disk spin-up. The frame in each of the panels is oriented perpendicular to the net angular momentum of all stars bound to the galaxy, thereby showing the edge-on view. The remnants of the protogalaxy are arranged in a close-to-spherical bulge-like formation. As we show in the companion paper \citep{semenov23b}, the formation of such a bulge is associated with the steepening of gravitational potential near the galaxy center. As was shown by \citet{hopkins23disk}, such a steepening can make infalling gas susceptible to disk formation via preferential dumping of radial orbits.

The bottom set of panels shows the stars that were formed between the time steps shown in the other two sets: from $\tspinup - 0.5$ Gyr to $\tspinup + 3$ Gyr. Newly formed stars are preferentially arranged in a disk configuration. This is the disk whose formation is reflected in the sharp increase of the median $\vrot$ of low-metallicity stars at redshift $z=0$ (i.e., the spin-up feature; see Figure~\ref{fig:FeH-vrot}). In some cases, the galaxy still undergoes mergers---e.g., the second and third panels in the first row and the second panel in the second row---but these stellar disks survive these mergers.

The stellar mass growth and the contribution from mergers during disk formation are quantified in Figure~\ref{fig:f-exsitu}. The evolution of different quantities is stacked among our early-spin-up galaxies, with the time offset to align them at the moment of disk spin-up ($\tspinup$; vertical dotted line). For reference, the top panel shows the median and $1\sigma$ scatter of $\vrot$ for young ($<100$ Myr) stars. Roughly 1 Gyr prior to $\tspinup$, a coherently rotating young stellar disk starts forming. By $\tspinup + 3$ Gyr, this disk acquires most of the final rotational velocity ($\sim 220 \kms$ for MW-like galaxies) and becomes rotation-dominated.

As the middle panel shows, both the total stellar mass in the main progenitor and the ex situ mass brought in by mergers rapidly increase early on before the disk forms. At later times, the total stellar mass continues increasing at a higher rate than the ex situ mass. From the bottom panel, the instantaneous ex situ mass fraction is the highest early on, $\sim 20\%\text{--}30\%$ when the disk starts forming. Although subdominant, such a mass fraction corresponds to quite significant mergers, with mass ratios between $\sim$1:4 and 1:2, assuming that most of the mass is delivered by the second-most-massive progenitor. Such major mergers can destroy disks formed at earlier stages (see, e.g., halo ID 523889 in Figure~\ref{fig:maps-proto}). From the analysis of our late-spin-up galaxies, we find that mergers with mass ratios as low as 1:10 can destroy pre-existing disks.

\begin{figure}
\centering
\includegraphics[width=\columnwidth]{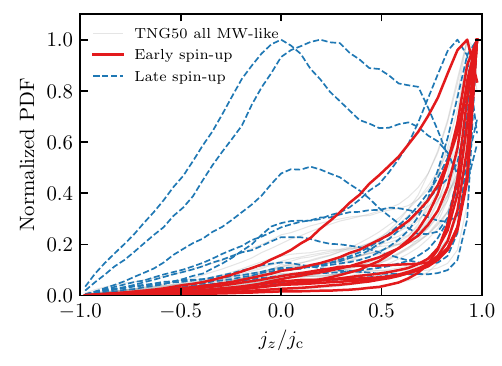}\\
\includegraphics[width=\columnwidth]{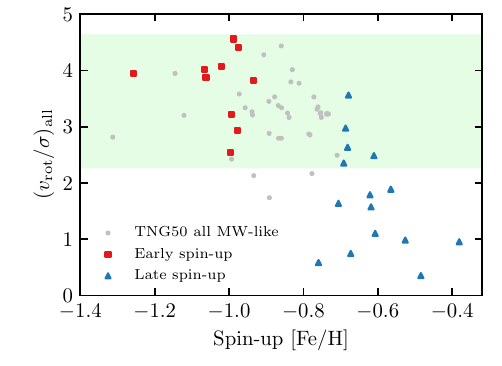}
\caption{\label{fig:diskiness} Early-spin-up galaxies are disk-dominated, while late-spin-up galaxies tend to have more pronounced spheroidal components. {\bf Top:} distribution of stellar orbital circularities ($j_z/j_{\rm c}$, see Section~\ref{sec:methods}) for the entire sample of MW-like TNG50 galaxies with highlighted early- and late-spin-up galaxies (solid red and dashed blue lines, respectively). In early-spin-up galaxies, circularities pile up toward $j_z/j_{\rm c} \sim 1$, corresponding to disk-like configurations, while many of the late-spin-up galaxies exhibit a prominent peak near $j_z/j_{\rm c} \sim 0$, corresponding to a spheroid. {\bf Bottom:} final rotational support of galaxies as a function of the stellar metallicity at which the spin-up occurs. \newtext{The ratio $\vsall$ is calculated using all star particles outside the central 3 kpc to exclude the bulge. The green shaded region indicates the typical values observed in nearby disk galaxies \citep{kregel05}.} Again, rotational support shows a prominent decreasing trend with the spin-up metallicity, indicating that earlier spin-up results in more prominent disks. }
\end{figure}

The results of this and the previous subsection suggest that early- and late-spin-up galaxies do not form through distinct channels, but are rather at the opposite ends of a continuous population shaped by the interplay between disk formation and survival through mergers. Early-spin-up galactic disks form at the end of the early active merger phase. In many cases, the formation of the disk is associated with the coalescence of comparable-mass galaxies, some of which could already develop a disk. Late-spin-up galaxies, on the other hand, assemble their final disk late, as a result of a merger destructing the pre-existing disk.

\subsection{Earlier-forming Disks Are More Disk-dominated}
\label{sec:results:diskiness}

The differences in the timing of disk formation described in previous subsections also translate into differences in morphological properties of galactic disks at redshift $z=0$, such as the mass fractions in the thin disk or the overall rotational support of the galaxy. Indeed, the earlier a gaseous disk can form, the more time there is to produce a massive stellar disk, as is evident from the visual inspection of some of the example galaxies (recall Figure~\ref{fig:maps}, which shows early- and late-spin-up galaxies in the top and bottom rows, respectively). This difference is quantified in Figure~\ref{fig:diskiness}.

The top panel of Figure~\ref{fig:diskiness} shows the distribution of orbital circularities, $j_z/j_{\rm c}$, of the star particles from the galactic disks (see Section~\ref{sec:methods:analysis} for the definition). For early-spin-up galaxies (solid red lines), the circularity piles up toward $j_z/j_{\rm c} \sim 1$, corresponding to a disk configuration with circular orbits. The late-spin-up galaxies (dashed blue lines), again, show a significant variation, with most of the galaxies exhibiting a broad peak near $j_z/j_{\rm c} \sim 0$, corresponding to a spheroidal configuration dominated by radial orbits. Some of the late-spin-up galaxies are, however, disk-dominated, similar to the early-spin-up sample.

The bottom panel of Figure~\ref{fig:diskiness} shows the relation between the metallicity at which the spin-up occurs and the rotational support of the galaxy, measured as the ratio of the median $\vrot$ to its scatter (half of the 16--84 interpercentile range) for all stars outside of the central 3 kpc.\footnote{The central 3 kpc were excluded to remove the contribution from the bulge.} Rotational support exhibits a clear decreasing trend with the spin-up [Fe/H], with early-spin-up galaxies (red squares) being rotationally supported, $\vsall \sim 2.5\text{--}5$ \newtext{in agreement with the typical values reported for nearby disk galaxies by \citet[][green shaded region]{kregel05},} while late-spin-up galaxies generally have smaller values of $\vsall \sim 0\text{--}3$, \newtext{implying the significant or even dominant contribution of the nondisk component}. 
Note that these values of $\vsall$ are lower than $\vsyoung = 6$ used to select our sample of MW-like galaxies (see Section~\ref{sec:methods:sample}) because the latter was applied only to stars younger than 100 Myr, while Figure~\ref{fig:diskiness} includes all ages. 

The trend shown in the figure qualitatively agrees with the findings of \citet{mccluskey23} that, in FIRE-2 simulations, disks that formed earlier exhibit stronger rotational support of the present-day young stellar disk. As these authors point out, such a trend may explain why the MW has a relatively high value of $\vsyoung \sim 8$ \newtext{ \citep[e.g.,][]{nordstrom04} } compared to, e.g., $\vsyoung \sim 3\text{--}5$ in M31 and M33 \newtext{ \citep[][]{dorman15,quirk22} }. Note, however, that the quantitative details of this trend can depend on star formation and AGN feedback modeling: even though FIRE-2 galaxies exhibit late spin-up features \citepalias[see][]{bk22}, none of them are bulge-dominated, as is the case for some of the TNG50 late-spin-up galaxies shown in the figure.

\section{Discussion}
\label{sec:discussion}

\subsection{Galactic Disk Formation Scenario}

The above results paint a rather straightforward picture of galaxy evolution that is reminiscent of the classical works, where the formation of galactic disks is driven by halo growth and gas cooling \citep[e.g.,][]{fall-efstathiou80,mo-mao-white98}. In this picture, the diversity of disk formation times for the galaxies that we see today is mainly driven by the differences in mass accretion histories. Some halos accumulate enough mass early and rapidly, leading to the early formation of disks, while others form over more extended timescales. In addition, significant mergers, if they occur, can destroy a pre-existing disk, resetting the final disk formation to a much later time \citep[see also][]{sotillo-ramos22}. 

\newtext{Recent galaxy formation simulations generally show that galactic disks form in dark matter halos above a mass threshold of $\Mvir \gtrsim 10^{11} \Msun$, corresponding to the galaxy stellar mass of $M_\star \gtrsim 10^9 \Msun$ \citep[e.g.,][]{el-badry18,pillepich19,dekel20,semenov23b}. In the companion paper \citep{semenov23b}, we show that in TNG50, this mass threshold is associated with the development of corotating circumgalactic medium feeding the disk with recycled outflows, bulge formation as a result of the final merger in the protogalaxy assembly, and the establishment of the hot gaseous halo. A strong bulge can provide a centrally concentrated steep potential around which infalling gas tends to settle in a disk \citep{hopkins23disk}, while the hot halo can transform the accretion mode, leading to smoother accretion of the inflowing gas with self-aligned angular momentum \citep{stern19,stern20,stern21,stern23}. The existence of such a mass threshold makes the disk formation naturally correlated with the halo mass assembly history, as the moment of disk spin-up becomes dependent on when a given halo crosses that threshold.}

In this sense, early- and late-spin-up galaxies are not different in the channels via which they form, but they rather occupy the opposite ends of the same galaxy population, shaped by the interplay of disk formation and destruction by mergers. It is therefore not surprising that early- and late-spin-up galaxies overlap in many of the properties that we have explored in this paper, such as the distribution of stellar $\vrot$ vs. age (the left panels of Figure~\ref{fig:FeH-vrot}), mass assembly history (Figure~\ref{fig:Mvir}), and different measures of diskiness at redshift $z=0$ (Figure~\ref{fig:diskiness}).

To produce an early-formed galactic disk with stellar chemokinematic signatures similar to our own MW, two conditions need to be satisfied: (i) its host halo must form early; and (ii) no destructive mergers can occur afterward.
This conclusion is consistent with previous theoretical and observational arguments that the MW did not experience significant mergers in the last $\sim 10$ Gyr \newtext{that would otherwise significantly perturb or destroy the disk} \citep[e.g.,][]{toth-ostriker92,wyse01,hammer07,belokurov20}. In addition, results from cosmological zoom-in simulations of MW-like galaxies suggest that producing a kinematic signature of a GSE-like merger biases the galaxy sample toward early mass assembly histories \citep{fattahi19,dillamore22}. Just like the case for the spin-up feature (Figure~\ref{fig:Mvir}), this result suggests a correlation between early mass assembly and GSE-like halo stars kinematics, but being independent, these two lines of evidence reinforce the case for the early assembly of the MW.

Importantly, our results in Section~\ref{sec:results:age-metallicity} also show that it is not enough to only form the disk early; the disk must form at low enough metallicities, which typically correspond to the earliest rapid enrichment stage. Consequently, only 1/4 of disks formed $>10$ Gyr ago exhibit spin-up features at metallicities [Fe/H] $\lesssim -1$ like the MW. 
This also implies that the population of metal-poor stars found in the MW's present-day protogalaxy remnant, or Aurora, imprints this early rapid enrichment stage \citep[see also][]{bk22,bk23,myeong22}.
Together, all these factors make the MW rather unusual, albeit consistent with the overall population of MW-mass disk galaxies at $\sim 10\%$ level.
One important implication of this result is that when a spin-up-like feature is measured in other MW-like galaxies (e.g., M31), it is likely to occur at higher metallicities than that in the MW.

\begin{figure}
\centering
\includegraphics[width=\columnwidth]{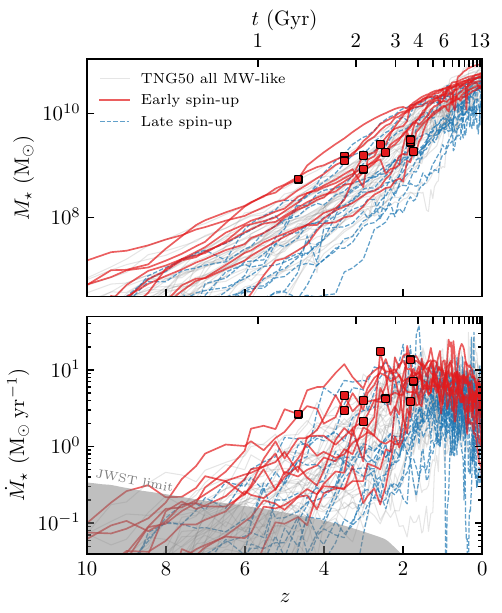}
\caption{\label{fig:history-z} Evolution of stellar masses (top) and SFRs (bottom) for the galaxies from our MW-mass TNG50 sample. Both $M_\star$ and $\SFR$ are computed within twice the radius containing half of the total SFR. The red squares show the moment of disk formation ($\tspinup$; see Section~\ref{sec:results:protogalaxy}). The shaded region in the bottom panel shows the rest-frame UV sensitivity of JWST in the F115W band computed with the Flexible Stellar Population Synthesis package \citep{fsps,conroy10-fsps,python-fsps} assuming 0.1 solar metallicity and a constant, 100 Myr long star formation burst. Early-spin-up MW analogs occupy the upper tail of the overall population and can be detected by JWST out to $z \sim 8$, well before their disks form.}
\end{figure}

Because of their early assembly, progenitors of chemokinematic analogs of the MW are significantly more massive and actively star-forming than an average progenitor of an MW-mass disk galaxy over their entire history, making them more easily detectable. Figure~\ref{fig:history-z} demonstrates this quantitatively, by showing stellar mass and SFR histories for our early- and late-spin-up galaxies. Both the $M_\star$ and SFR of early-spin-up MW analogs occupy the upper envelope of the overall population and reach up to an order of magnitude higher values than the population average. For reference, the red squares indicate the moment of disk spin-up and the shaded region in the lower panel shows the rest-frame UV sensitivity of JWST. JWST can detect progenitors of MW analogs out to $z \sim 8$ and provide a view of their disk formation at $z \sim 5\text{--}2$.

Despite the fact that, in simulations, the most massive main progenitor of the MW can be tracked out to this early epoch, this protogalaxy stage is associated with active mergers that bring in a significant fraction of mass corresponding to $\sim 1:2\text{--}1:4$ merger mass ratios (Section~\ref{sec:results:protogalaxy}). In the context of galactic archeology, such major mergers can complicate the identification of a single early progenitor in the present-day stellar distribution using chemical abundance cuts. This will be the case if the major pieces of the protogalaxy are evolving in similar ways prior to the coalescence, which will result in similar abundance patterns, making it hard to distinguish this scenario from the in situ growth of a single well-defined progenitor. We leave the investigation of this uncertainty using cosmological simulations to a future study.

\subsection{Delayed Disk Formation in Zoom-in Simulations}

Our results can resolve the problem pointed out by \citetalias{bk22} that, in zoom-in galaxy formation simulations of MW-like galaxies (FIRE and Auriga), the spin-up feature occurs at significantly higher stellar metallicities ([Fe/H] $\sim -0.5$) than observed in the MW ([Fe/H] $\lesssim -1$; see Figure~17 in \citetalias{bk22}). From the enrichment histories presented in the same paper, such an offset implies that the disk formation in these simulations is delayed by $\sim 2$ Gyr.

Our results show that, on average, the MW-mass galactic disks indeed spin up at later times and higher metallicities than the local MW data suggest. The formation history of the MW is unusual and, unless the initial conditions (ICs) of zoom-in simulations are selected to approximate such a history, the spin-up is likely to occur late. Neither FIRE nor Auriga specifically selects galaxies with MW-like assembly histories and therefore it is not surprising that they show the spin-up feature at high metallicities.
On the other hand, HESTIA simulations that were tailored to produce Local Group-like environments produce early-spin-up features for some of their MW and M31 analogs, although the scatter is large (in agreement with our findings) and half of the realizations result in a late spin-up at $-0.7 \lesssim \text{[Fe/H]} \lesssim -0.5$ \citep[see Figure~14 in][]{khoperskov22c}.

Interestingly, the spin-up features in the zoom-in simulations explored by \citetalias{bk22} are all clustered around [Fe/H] $\sim -0.5$, which is higher than the median value that we find in TNG50 and closer to the late-spin-up galaxies (see Figure~\ref{fig:FeH-vrot}). One reason for such apparently late spin-up is the difference in how the metallicity values are corrected in simulations. \citetalias{bk22} shift the metallicities of all stellar particles to match the median value in the MW data. As shown in Appendix~\ref{app:metallicity}, such a method indeed results in a later spin-up than our fiducial abundance-matching-based calibration (see Figure~\ref{fig:FeH-calibration}).
Another potential reason for the late disk formation in zoom-in runs is a bias introduced by the selection of the ICs. In some cases, the ICs have been specifically selected so that the galaxy forms late and spends less time in the highly nonlinear regime, making such simulations more computationally affordable and tractable for parameter studies. Such biases can result in late disk formation compared to the MW data.

\begin{figure}
\centering
\includegraphics[width=\columnwidth]{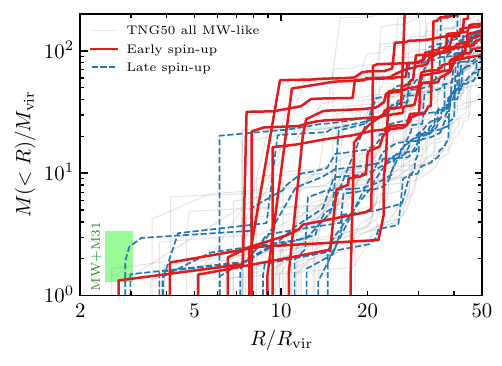}
\caption{\label{fig:environment} Comparison of the local environments for the sample of MW-like TNG galaxies. The $y$-axis shows a cumulative mass in all halos more massive than $0.3\;\Mvir$ of the galaxy host halo within the radius normalized by the virial radius of that host halo. The green square shows the range occupied by the MW and M31 pair, assuming $\Mvir = (1.3 \pm 0.3) \times 10^{12}\Msun$ and $\Rvir = 282 \pm 30\kpc$ for the MW \citep{mw-review}, $M_{\rm vir,M31} = (1.4 \pm 0.3) \times 10^{12}\Msun$ for M31 \citep{watkins10} and $R \approx 765\kpc$ for the distance between the two. The simulation results do not show any clear difference in the environments between early- and late-spin-up galaxies within $R < 20 \Rvir$, while on larger scales early-spin-up ones tend to reside in stronger overdensities.}
\end{figure}

A remarkable case of a FIRE galaxy that appears to form a disk early is Romeo, one of the galaxies from the suite of Local Group analog simulations with the FIRE model, ELVIS \citep{elvis}. This galaxy was not investigated in \citetalias{bk22}, but the results of \citet{yu22} and \citet{mccluskey23} suggest that it forms a thick disk at lookback $t \sim 10\text{--}11$ Gyr. This result can be a consequence of Local Group analogs assembling their mass earlier than isolated MW-mass galaxies \citep[e.g.,][]{santistevan20}, indicating that disks may preferentially form early in galaxy pairs and groups. Such environments correspond to stronger perturbations in the initial density field and their members likely have not experienced recent significant mergers, as a major one is yet to occur.

In Figure~\ref{fig:environment} we investigate this possibility by comparing the local environments of early- and late-spin-up galaxies in TNG50. This plot shows the cumulative distribution of the total mass in halos more massive than $0.3\;\Mvir$ of the host mass of a given galaxy as a function of the distance from that galaxy normalized by its $\Rvir$. Isolated galaxies populate the lower envelope of the distribution, while galaxies with a massive neighbor in their proximity correspond to the upper envelope. As the figure shows, there is no clear difference between the two samples up to $\sim 20 \Rvir$, implying that the immediate environments of early- and late-spin-up galaxies are not significantly different. Interestingly, at larger distances, the environments around early-spin-up galaxies tend to contain more mass in total, suggesting that the large-scale environment may affect how early galactic disks form.

\section{Summary and Conclusions}
\label{sec:summary}

Using the large-volume cosmological simulation TNG50 \citep{pillepich19,nelson19}, we have investigated the formation and evolution of MW-like galaxies in view of recent ``galactic archeology'' findings showing that the disk of our own MW formed early, within the first few Gyr of the Universe's evolution \citep{bk22,bk23,conroy22,rix22,xiang-rix22}. We specifically focused on the relation between the azimuthal velocities of stars and their metallicities, which imprints the emergence of the disk, or ``spin-up'', at metallicities $-1.3 \lesssim \text{[Fe/H]} \lesssim -1$ \citep{bk22}. As these authors showed, a subset of the FIRE and Auriga zoom-in cosmological simulations appear to be in tension with this result, predicting disk formation at 3--10 times higher stellar metallicities, implying a few Gyr delay.

Using a representative sample of MW-like galaxies from TNG50, we investigate the origin of this discrepancy and, more broadly, galactic disk formation in the early Universe. Our key findings can be summarized as follows.

\begin{enumerate}
    \item MW-like TNG50 galaxies all show a sharp spin-up feature with a significant variation of the metallicity at which this feature occurs, $-1.2 \lesssim \text{[Fe/H]} \lesssim -0.5$. The MW is not a typical MW-mass galaxy, but it is broadly consistent with this population, being at the low-metallicity end of the distribution within $\sim 2 \sigma$ from the median (see Figure~\ref{fig:FeH-vrot-median}). Such early-spin-up galaxies like our MW form disks early ($\gtrsim 10$ Gyr ago) and rapidly (on the timescale of $\sim 1$ Gyr), while late-spin-up galaxies show significant variation both in the timing and duration of the disk formation (Figure~\ref{fig:FeH-vrot}).

    \item The shape and scatter of the age--metallicity relation plays an important role in the formation of early- vs. late-spin-up galaxies. Early-spin-up disks form during the earliest stage of galaxy assembly associated with rapid chemical enrichment, while late-spin-up ones form at a later, slow enrichment stage (Figure~\ref{fig:spinup-age}). As a result, while $\sim 40\%$ of disks form $\gtrsim 10$ Gyr ago, only $\sim$1/4 of such disks have the spin-up feature at metallicities as low as in the MW, [Fe/H] $\lesssim -1$. In addition, this difference leads to late-spin-up galaxies exhibiting a sharper spin-up feature in metallicity and a broader variation of stellar ages at which this feature occurs (the top right panel of Figure~\ref{fig:FeH-vrot}).    

    \item The variation of the metallicity and timing of disk formation is mainly associated with the differences in the mass assembly history of galaxies. Early-spin-up galaxies assemble most of their mass early ($\gtrsim 10$ Gyr ago) and grow steadily after that (Figure~\ref{fig:Mvir}). In contrast, late-spin-up galaxies generally assemble over longer timescales and can experience significant mergers that can destroy a pre-existing disk, resetting the spin-up feature to the time and metallicity of the last destructive merger (Figure~\ref{fig:early-late-examples}).

    \item The pre-disk epoch of early-spin-up galaxy evolution is associated with active galaxy mergers. The remnant of this protogalaxy stage arranges a bulge-like structure around which the stellar disk is formed (Figure~\ref{fig:maps-proto}). These early mergers account for $\sim 20\%\text{--}30\%$ of the total stellar mass growth during disk formation (Figure~\ref{fig:f-exsitu}).

    \item The differences in the timing of the disk formation translate into the differences in the disk properties at redshift $z=0$. Early-spin-up galaxies generally have a higher fraction of stars in circular orbits and exhibit stronger rotational support of the stellar disk than late-spin-up galaxies (Figure~\ref{fig:diskiness}).
\end{enumerate}

Our results indicate that, while the MW is consistent with the overall population of MW-mass disk galaxies in TNG50, it is rather unusual, as it results from a coincidence of two factors: (i) the disk must form very early, during the epoch of rapid chemical enrichment $> 10$ Gyr ago; and (ii) it must not encounter destructive mergers in the past 10 Gyr. 
We find that only $\sim 10\%$ of all MW-mass disk galaxies are both early-forming and metal-poor during disk formation, in agreement with the MW data. This result implies that spin-up-like features in other MW-like galaxies (e.g., M31) are more likely to occur at higher metallicities than in the MW. Therefore, future measurements of such kinematic signatures of disk formation in nearby galaxies would provide significant constraints on theoretical models of galaxy assembly.

Overall, early- and late-spin-up galaxies are not formed through different channels, but rather occupy the opposite tails of the same galaxy population shaped by the interplay between the disk formation and its survival through mergers. In this paper, we have mainly investigated the effects of galaxy assembly history on the stellar chemokinematic signatures of disk formation. In a companion paper \citep{semenov23b}, we focus on the physical factors driving disk formation in the early Universe.

\section*{Acknowledgements}
We thank the anonymous referee for their constructive feedback that improved the manuscript.
We are deeply grateful to Vasily Belokurov, Andrey Kravtsov, and Jonathan Stern for their comprehensive comments on the draft that improved this paper.
We also thank Greg Bryan, Claude-Andr\'e Faucher-Gigu\`ere, Drummond Fielding, Shy Genel, Alexander Gurvich, Zachary Hafen, Philip Hopkins, Andrew Wetzel, and the participants of the Disk Formation Workshop at UC Irvine for inspirational and insightful discussions.
We also thank Benjamin Johnson for his help with deriving JWST sensitivity limits.
The analyses presented in this paper were performed on the FASRC Cannon cluster supported by the FAS Division of Science Research Computing Group at Harvard University and on the HPC system Vera at the Max Planck Computing and Data Facility, and we thank Annalisa Pillepich for providing access to Vera.
Support for V.S. was provided by NASA through the NASA Hubble Fellowship grant HST-HF2-51445.001-A awarded by the Space Telescope Science Institute, which is operated by the Association of Universities for Research in Astronomy, Inc., for NASA, under contract NAS5-26555, and by Harvard University through the Institute for Theory and Computation Fellowship. 
D.N. acknowledges funding from the Deutsche Forschungsgemeinschaft (DFG) through an Emmy Noether Research Group (grant number NE 2441/1-1).
The analyses presented in this paper were greatly aided by the following free software packages: {\tt NumPy} \citep{numpy_ndarray}, {\tt SciPy} \citep{scipy}, and {\tt Matplotlib} \citep{matplotlib}. We have also used the Astrophysics Data Service (\href{http://adsabs.harvard.edu/abstract_service.html}{\tt ADS}) and \href{https://arxiv.org}{\tt arXiv} preprint repository extensively during this project and the writing of the paper.

\appendix

\section{Metallicity Calibration}
\label{app:metallicity}

\begin{figure*}
\centering
\includegraphics[width=\textwidth]{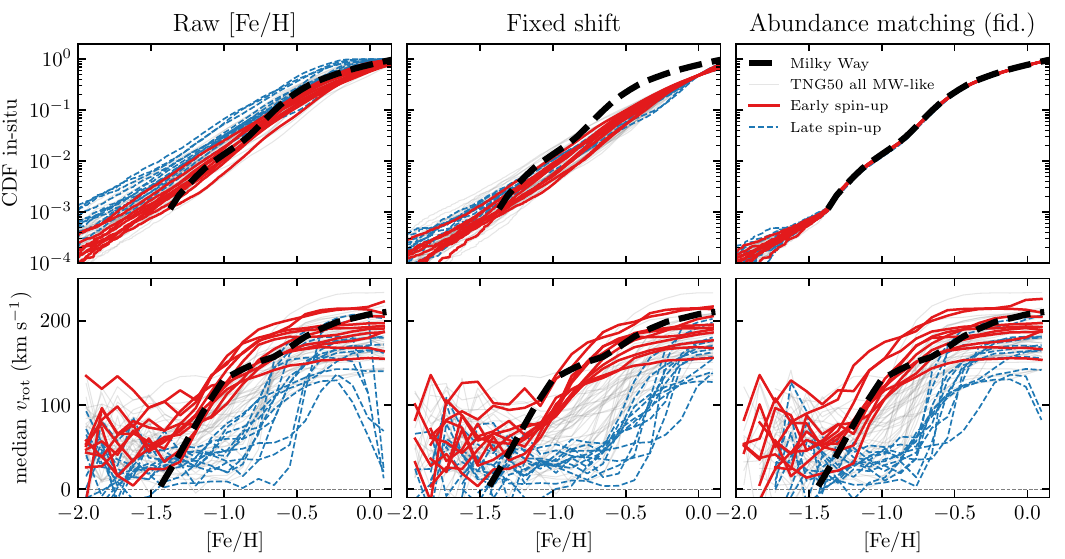}\\
\includegraphics[width=\textwidth]{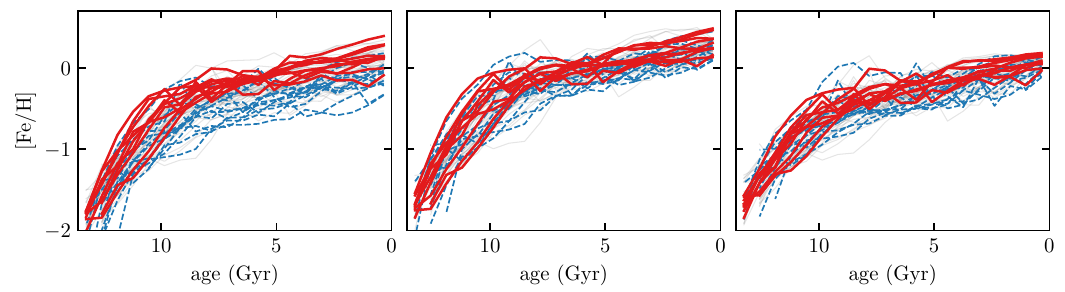}
\caption{\label{fig:FeH-calibration} Effect of star particle metallicity rescaling on the cumulative distribution of stellar [Fe/H] (top row), the relation between median $\vrot$ and [Fe/H] (middle row), and the age--metallicity relation (bottom row). The left column shows the results for the ``raw'' values of [Fe/H] taken directly from the final TNG50 snapshot. The middle column shows the results when the [Fe/H] of stars from each galaxy are shifted by a constant factor, such that the median [Fe/H] becomes 0 (this is the method used by \citetalias{bk22} in their analysis of simulation results). Finally, the last column shows our fiducial method based on abundance matching of star particles, where the order of [Fe/H] is preserved, but its value for each particle is scaled, such that the final CDF matches that from the observational sample of \citetalias{bk22} shown with the black dashed line. The choice of [Fe/H] calibration affects both the scatter and steepness of the spin-up feature in the [Fe/H]--$\vrot$ plane and the scatter and shape of the age--metallicity relation.}
\end{figure*}

\begin{figure}
\centering
\includegraphics[width=\columnwidth]{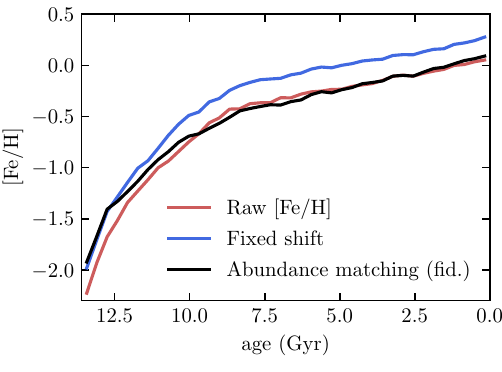}
\caption{\label{fig:age-FeH-median} Comparison of the median age--metallicity relations from the bottom row of Figure~\ref{fig:FeH-calibration}.}
\end{figure}

Assumptions about metal yields, stellar evolution models, and the IMF constitute key uncertainties for predicting the metallicity distribution and its evolution in galaxy formation simulations. To mitigate these uncertainties, we rescale the metallicities of the star particles predicted in TNG50, such that the new distribution better approximates the observed one. Figure~\ref{fig:FeH-calibration} shows how the resulting relation between the median $\vrot$ and [Fe/H] and the age--metallicity relation change depending on the choice of this rescaling procedure. In particular, we show the results for three choices:

\begin{itemize}
    \item \emph{Raw [Fe/H] values.} We use the ``raw'' values of [Fe/H] predicted in TNG50 for each star particle, without applying any correction. As the top left panel in Figure~\ref{fig:FeH-calibration} shows, the CDFs of raw [Fe/H] are significantly flatter than that reported for the MW by \citetalias{bk22} (dashed black line), implying an overabundance of metal-poor stars and slow enrichment on average.

    \item \emph{Fixed shift.} All values of [Fe/H] are shifted by the same constant factor, such that the median value of [Fe/H] over all stars within the spatial selection region ($R = 5\text{--}11\kpc$ and $|z|<3\kpc$; see Section~\ref{sec:methods:sample}) becomes 0. This constant factor differs from galaxy to galaxy. As a result, all CDFs intersect at [Fe/H] $=0$ and CDF $=0.5$, while their shapes remain flat, as they are invariant to the shift by a constant factor. This is the method adopted by \citetalias{bk22} in their analysis of zoom-in simulation results.

    \item \emph{Abundance matching.} The original order of ``raw'' [Fe/H] is preserved, but the individual values of [Fe/H] are shifted, such that the resulting cumulative abundance of [Fe/H] matches the observed CDF. At low metallicities, where the \citetalias{bk22} sample runs out of stars, [Fe/H] $\lesssim -1.3$, we apply a constant shift, such that all simulated CDFs converge at the observed CDF. We apply this method only using the star particles from the same spatial region that we use in comparison with the \citetalias{bk22} results ($R = 5\text{--}11\kpc$ and $|z|<3\kpc$; see Section~\ref{sec:methods:sample}). This is our fiducial method that we use in the rest of the paper, as it is designed to match the observed CDF exactly.
\end{itemize}

The middle row in Figure~\ref{fig:FeH-calibration} shows the effect of the above choice on the location and sharpness of the spin-up feature in the [Fe/H]--$\vrot$ plane. 
\newtext{The abundance matching of stellar [Fe/H] shifts the spin-up feature to a $\sim0.1\text{--}0.2$ dex higher metallicity and makes this feature significantly sharper. The net metallicity offset is caused by the increase of [Fe/H] for the low-metallicity stars, suggesting more rapid early enrichment than the raw TNG50 results imply.}
The spin-up becomes sharper because the abundance-matched CDF is significantly steeper, which means that the particle-to-particle variation of metallicities in one galaxy is smaller. Such a smaller variation implies that metallicities change more gradually as a function of time (stellar age) and, as a result, the formation of a disk over a fixed timescale translates into a sharper change of $\vrot$ as a function of metallicity. This is also the reason why this feature is sharper for late-spin-up galaxies than for early-spin-up ones: it is because the increase of [Fe/H] becomes more gradual at later times.

The middle row in Figure~\ref{fig:FeH-calibration} shows the effect of [Fe/H] rescaling on the age--metallicity relation. The early- and late-spin-up samples are clearly separated when raw [Fe/H] values are used, with the latter ones preferentially having lower [Fe/H] at a fixed age. Applying either a fixed shift or abundance matching reduces the galaxy-to-galaxy variation of the relation, and the separation between the subsamples becomes less apparent.

The median age--metallicity relations for these different calibrations are compared in Figure~\ref{fig:age-FeH-median}. The overall shape of the relation is not very sensitive to the choice of calibration, except for the constant offset of $\sim 0.2$ dex for the ``fixed shift'' method, which corresponds to the difference between the median value of raw [Fe/H] and that imposed by the method. This constant offset can be one of the reasons why, in the analysis of \citepalias{bk22}, the FIRE and Auriga simulations show the spin-up feature at somewhat higher metallicities than the median value in TNG50 (see Figure~\ref{fig:FeH-vrot-median}).

\bibliographystyle{aasjournal}
\bibliography{}

\begin{thebibliography}{}
\expandafter\ifx\csname natexlab\endcsname\relax\def\natexlab#1{#1}\fi
\providecommand{\url}[1]{\href{#1}{#1}}
\providecommand{\dodoi}[1]{doi:~\href{http://doi.org/#1}{\nolinkurl{#1}}}
\providecommand{\doeprint}[1]{\href{http://ascl.net/#1}{\nolinkurl{http://ascl.net/#1}}}
\providecommand{\doarXiv}[1]{\href{https://arxiv.org/abs/#1}{\nolinkurl{https://arxiv.org/abs/#1}}}

\bibitem[{{Agertz} \& {Kravtsov}(2015)}]{agertz-kravtsov15}
{Agertz}, O., \& {Kravtsov}, A.~V. 2015, \apj, 804, 18,
  \dodoi{10.1088/0004-637X/804/1/18}

\bibitem[{{Agertz} \& {Kravtsov}(2016)}]{agertz-kravtsov16}
---. 2016, \apj, 824, 79, \dodoi{10.3847/0004-637X/824/2/79}

\bibitem[{{Agertz} {et~al.}(2021){Agertz}, {Renaud}, {Feltzing}, {Read},
  {Ryde}, {Andersson}, {Rey}, {Bensby}, \& {Feuillet}}]{vintergatan1}
{Agertz}, O., {Renaud}, F., {Feltzing}, S., {et~al.} 2021, \mnras, 503, 5826,
  \dodoi{10.1093/mnras/stab322}

\bibitem[{{Belokurov} \& {Kravtsov}(2022)}]{bk22}
{Belokurov}, V., \& {Kravtsov}, A. 2022, \mnras, 514, 689,
  \dodoi{10.1093/mnras/stac1267}

\bibitem[{{Belokurov} \& {Kravtsov}(2023)}]{bk23}
---. 2023, \mnras, 525, 4456, \dodoi{10.1093/mnras/stad2241}

\bibitem[{{Belokurov} {et~al.}(2020){Belokurov}, {Sanders}, {Fattahi}, {Smith},
  {Deason}, {Evans}, \& {Grand}}]{belokurov20}
{Belokurov}, V., {Sanders}, J.~L., {Fattahi}, A., {et~al.} 2020, \mnras, 494,
  3880, \dodoi{10.1093/mnras/staa876}

\bibitem[{{Bird} {et~al.}(2013){Bird}, {Kazantzidis}, {Weinberg}, {Guedes},
  {Callegari}, {Mayer}, \& {Madau}}]{bird13}
{Bird}, J.~C., {Kazantzidis}, S., {Weinberg}, D.~H., {et~al.} 2013, \apj, 773,
  43, \dodoi{10.1088/0004-637X/773/1/43}

\bibitem[{{Bird} {et~al.}(2021){Bird}, {Loebman}, {Weinberg}, {Brooks},
  {Quinn}, \& {Christensen}}]{bird21}
{Bird}, J.~C., {Loebman}, S.~R., {Weinberg}, D.~H., {et~al.} 2021, \mnras, 503,
  1815, \dodoi{10.1093/mnras/stab289}

\bibitem[{{Bland-Hawthorn} \& {Gerhard}(2016)}]{mw-review}
{Bland-Hawthorn}, J., \& {Gerhard}, O. 2016, \araa, 54, 529,
  \dodoi{10.1146/annurev-astro-081915-023441}

\bibitem[{{Bonaca} {et~al.}(2020){Bonaca}, {Conroy}, {Cargile}, {Naidu},
  {Johnson}, {Zaritsky}, {Ting}, {Caldwell}, {Han}, \& {van Dokkum}}]{bonaca20}
{Bonaca}, A., {Conroy}, C., {Cargile}, P.~A., {et~al.} 2020, \apjl, 897, L18,
  \dodoi{10.3847/2041-8213/ab9caa}

\bibitem[{{Bournaud} {et~al.}(2009){Bournaud}, {Elmegreen}, \&
  {Martig}}]{bournaud09}
{Bournaud}, F., {Elmegreen}, B.~G., \& {Martig}, M. 2009, \apjl, 707, L1,
  \dodoi{10.1088/0004-637X/707/1/L1}

\bibitem[{{Brook} {et~al.}(2004){Brook}, {Kawata}, {Gibson}, \&
  {Freeman}}]{brook04}
{Brook}, C.~B., {Kawata}, D., {Gibson}, B.~K., \& {Freeman}, K.~C. 2004, \apj,
  612, 894, \dodoi{10.1086/422709}

\bibitem[{{Brook} {et~al.}(2012){Brook}, {Stinson}, {Gibson}, {Kawata},
  {House}, {Miranda}, {Macci{\`o}}, {Pilkington}, {Ro{\v{s}}kar}, {Wadsley}, \&
  {Quinn}}]{brook12}
{Brook}, C.~B., {Stinson}, G.~S., {Gibson}, B.~K., {et~al.} 2012, \mnras, 426,
  690, \dodoi{10.1111/j.1365-2966.2012.21738.x}

\bibitem[{{Brooks} {et~al.}(2009){Brooks}, {Governato}, {Quinn}, {Brook}, \&
  {Wadsley}}]{brooks09}
{Brooks}, A.~M., {Governato}, F., {Quinn}, T., {Brook}, C.~B., \& {Wadsley}, J.
  2009, \apj, 694, 396, \dodoi{10.1088/0004-637X/694/1/396}

\bibitem[{{Bryan} \& {Norman}(1998)}]{bryan-norman98}
{Bryan}, G.~L., \& {Norman}, M.~L. 1998, \apj, 495, 80, \dodoi{10.1086/305262}

\bibitem[{{Chabrier}(2003)}]{chabrier03}
{Chabrier}, G. 2003, \pasp, 115, 763, \dodoi{10.1086/376392}

\bibitem[{{Clauwens} {et~al.}(2018){Clauwens}, {Schaye}, {Franx}, \&
  {Bower}}]{clauwens18}
{Clauwens}, B., {Schaye}, J., {Franx}, M., \& {Bower}, R.~G. 2018, \mnras, 478,
  3994, \dodoi{10.1093/mnras/sty1229}

\bibitem[{{Conroy} \& {Gunn}(2010)}]{conroy10-fsps}
{Conroy}, C., \& {Gunn}, J.~E. 2010, \apj, 712, 833,
  \dodoi{10.1088/0004-637X/712/2/833}

\bibitem[{{Conroy} {et~al.}(2009){Conroy}, {Gunn}, \& {White}}]{fsps}
{Conroy}, C., {Gunn}, J.~E., \& {White}, M. 2009, \apj, 699, 486,
  \dodoi{10.1088/0004-637X/699/1/486}

\bibitem[{{Conroy} {et~al.}(2022){Conroy}, {Weinberg}, {Naidu}, {Buck},
  {Johnson}, {Cargile}, {Bonaca}, {Caldwell}, {Chandra}, {Han}, {Johnson},
  {Speagle}, {Ting}, {Woody}, \& {Zaritsky}}]{conroy22}
{Conroy}, C., {Weinberg}, D.~H., {Naidu}, R.~P., {et~al.} 2022, arXiv e-prints,
  arXiv:2204.02989.
\newblock \doarXiv{2204.02989}

\bibitem[{{Conselice}(2014)}]{conselice14}
{Conselice}, C.~J. 2014, \araa, 52, 291,
  \dodoi{10.1146/annurev-astro-081913-040037}

\bibitem[{{Conselice} {et~al.}(2008){Conselice}, {Rajgor}, \&
  {Myers}}]{conselice08}
{Conselice}, C.~J., {Rajgor}, S., \& {Myers}, R. 2008, \mnras, 386, 909,
  \dodoi{10.1111/j.1365-2966.2008.13069.x}

\bibitem[{{Correa Magnus} \& {Vasiliev}(2022)}]{correamagnus22}
{Correa Magnus}, L., \& {Vasiliev}, E. 2022, \mnras, 511, 2610,
  \dodoi{10.1093/mnras/stab3726}

\bibitem[{{Davison} {et~al.}(2020){Davison}, {Norris}, {Pfeffer}, {Davies}, \&
  {Crain}}]{davison20}
{Davison}, T.~A., {Norris}, M.~A., {Pfeffer}, J.~L., {Davies}, J.~J., \&
  {Crain}, R.~A. 2020, \mnras, 497, 81, \dodoi{10.1093/mnras/staa1816}

\bibitem[{{Deason} {et~al.}(2021){Deason}, {Erkal}, {Belokurov}, {Fattahi},
  {G{\'o}mez}, {Grand}, {Pakmor}, {Xue}, {Liu}, {Yang}, {Zhang}, \&
  {Zhao}}]{deason21}
{Deason}, A.~J., {Erkal}, D., {Belokurov}, V., {et~al.} 2021, \mnras, 501,
  5964, \dodoi{10.1093/mnras/staa3984}

\bibitem[{{Dekel} \& {Birnboim}(2006)}]{dekel-birnboim06}
{Dekel}, A., \& {Birnboim}, Y. 2006, \mnras, 368, 2,
  \dodoi{10.1111/j.1365-2966.2006.10145.x}

\bibitem[{{Dekel} {et~al.}(2020){Dekel}, {Ginzburg}, {Jiang}, {Freundlich},
  {Lapiner}, {Ceverino}, \& {Primack}}]{dekel20}
{Dekel}, A., {Ginzburg}, O., {Jiang}, F., {et~al.} 2020, \mnras, 493, 4126,
  \dodoi{10.1093/mnras/staa470}

\bibitem[{{Dillamore} {et~al.}(2022){Dillamore}, {Belokurov}, {Font}, \&
  {McCarthy}}]{dillamore22}
{Dillamore}, A.~M., {Belokurov}, V., {Font}, A.~S., \& {McCarthy}, I.~G. 2022,
  \mnras, 513, 1867, \dodoi{10.1093/mnras/stac1038}

\bibitem[{{Dorman} {et~al.}(2015){Dorman}, {Guhathakurta}, {Seth}, {Weisz},
  {Bell}, {Dalcanton}, {Gilbert}, {Hamren}, {Lewis}, {Skillman}, {Toloba}, \&
  {Williams}}]{dorman15}
{Dorman}, C.~E., {Guhathakurta}, P., {Seth}, A.~C., {et~al.} 2015, \apj, 803,
  24, \dodoi{10.1088/0004-637X/803/1/24}

\bibitem[{{El-Badry} {et~al.}(2018){El-Badry}, {Quataert}, {Wetzel}, {Hopkins},
  {Weisz}, {Chan}, {Fitts}, {Boylan-Kolchin}, {Kere{\v{s}}},
  {Faucher-Gigu{\`e}re}, \& {Garrison-Kimmel}}]{el-badry18}
{El-Badry}, K., {Quataert}, E., {Wetzel}, A., {et~al.} 2018, \mnras, 473, 1930,
  \dodoi{10.1093/mnras/stx2482}

\bibitem[{{Elmegreen} {et~al.}(2007){Elmegreen}, {Elmegreen}, {Ravindranath},
  \& {Coe}}]{elmegreen07}
{Elmegreen}, D.~M., {Elmegreen}, B.~G., {Ravindranath}, S., \& {Coe}, D.~A.
  2007, \apj, 658, 763, \dodoi{10.1086/511667}

\bibitem[{{Fall} \& {Efstathiou}(1980)}]{fall-efstathiou80}
{Fall}, S.~M., \& {Efstathiou}, G. 1980, \mnras, 193, 189,
  \dodoi{10.1093/mnras/193.2.189}

\bibitem[{{Fattahi} {et~al.}(2019){Fattahi}, {Belokurov}, {Deason}, {Frenk},
  {G{\'o}mez}, {Grand}, {Marinacci}, {Pakmor}, \& {Springel}}]{fattahi19}
{Fattahi}, A., {Belokurov}, V., {Deason}, A.~J., {et~al.} 2019, \mnras, 484,
  4471, \dodoi{10.1093/mnras/stz159}

\bibitem[{{Faucher-Gigu{\`e}re} {et~al.}(2009){Faucher-Gigu{\`e}re}, {Lidz},
  {Zaldarriaga}, \& {Hernquist}}]{faucher-giguere09}
{Faucher-Gigu{\`e}re}, C.-A., {Lidz}, A., {Zaldarriaga}, M., \& {Hernquist}, L.
  2009, \apj, 703, 1416, \dodoi{10.1088/0004-637X/703/2/1416}

\bibitem[{{Ferreira} {et~al.}(2022){Ferreira}, {Adams}, {Conselice},
  {Sazonova}, {Austin}, {Caruana}, {Ferrari}, {Verma}, {Trussler},
  {Broadhurst}, {Diego}, {Frye}, {Pascale}, {Wilkins}, {Windhorst}, \&
  {Zitrin}}]{ferreira22a}
{Ferreira}, L., {Adams}, N., {Conselice}, C.~J., {et~al.} 2022, \apjl, 938, L2,
  \dodoi{10.3847/2041-8213/ac947c}

\bibitem[{{Ferreira} {et~al.}(2023){Ferreira}, {Conselice}, {Sazonova},
  {Ferrari}, {Caruana}, {Tohill}, {Lucatelli}, {Adams}, {Irodotou}, {Marshall},
  {Roper}, {Lovell}, {Verma}, {Austin}, {Trussler}, \& {Wilkins}}]{ferreira22b}
{Ferreira}, L., {Conselice}, C.~J., {Sazonova}, E., {et~al.} 2023, \apj, 955,
  94, \dodoi{10.3847/1538-4357/acec76}

\bibitem[{{Foreman-Mackey} {et~al.}(2014){Foreman-Mackey}, {Sick}, \&
  {Johnson}}]{python-fsps}
{Foreman-Mackey}, D., {Sick}, J., \& {Johnson}, B. 2014, {python-fsps: Python
  bindings to FSPS (v0.1.1)}, v0.1.1, Zenodo,  Zenodo,
  \dodoi{10.5281/zenodo.12157}

\bibitem[{{F{\"o}rster Schreiber} \& {Wuyts}(2020)}]{forsterschreiber-wuyts20}
{F{\"o}rster Schreiber}, N.~M., \& {Wuyts}, S. 2020, \araa, 58, 661,
  \dodoi{10.1146/annurev-astro-032620-021910}

\bibitem[{{F{\"o}rster Schreiber} {et~al.}(2006){F{\"o}rster Schreiber},
  {Genzel}, {Lehnert}, {Bouch{\'e}}, {Verma}, {Erb}, {Shapley}, {Steidel},
  {Davies}, {Lutz}, {Nesvadba}, {Tacconi}, {Eisenhauer}, {Abuter}, {Gilbert},
  {Gillessen}, \& {Sternberg}}]{forsterschreiber06}
{F{\"o}rster Schreiber}, N.~M., {Genzel}, R., {Lehnert}, M.~D., {et~al.} 2006,
  \apj, 645, 1062, \dodoi{10.1086/504403}

\bibitem[{{F{\"o}rster Schreiber} {et~al.}(2009){F{\"o}rster Schreiber},
  {Genzel}, {Bouch{\'e}}, {Cresci}, {Davies}, {Buschkamp}, {Shapiro},
  {Tacconi}, {Hicks}, {Genel}, {Shapley}, {Erb}, {Steidel}, {Lutz},
  {Eisenhauer}, {Gillessen}, {Sternberg}, {Renzini}, {Cimatti}, {Daddi},
  {Kurk}, {Lilly}, {Kong}, {Lehnert}, {Nesvadba}, {Verma}, {McCracken},
  {Arimoto}, {Mignoli}, \& {Onodera}}]{forsterschreiber09}
{F{\"o}rster Schreiber}, N.~M., {Genzel}, R., {Bouch{\'e}}, N., {et~al.} 2009,
  \apj, 706, 1364, \dodoi{10.1088/0004-637X/706/2/1364}

\bibitem[{{F{\"o}rster Schreiber} {et~al.}(2011){F{\"o}rster Schreiber},
  {Shapley}, {Genzel}, {Bouch{\'e}}, {Cresci}, {Davies}, {Erb}, {Genel},
  {Lutz}, {Newman}, {Shapiro}, {Steidel}, {Sternberg}, \&
  {Tacconi}}]{forsterschreiber11}
{F{\"o}rster Schreiber}, N.~M., {Shapley}, A.~E., {Genzel}, R., {et~al.} 2011,
  \apj, 739, 45, \dodoi{10.1088/0004-637X/739/1/45}

\bibitem[{{Fraternali} {et~al.}(2021){Fraternali}, {Karim}, {Magnelli},
  {G{\'o}mez-Guijarro}, {Jim{\'e}nez-Andrade}, \& {Posses}}]{fraternali21}
{Fraternali}, F., {Karim}, A., {Magnelli}, B., {et~al.} 2021, \aap, 647, A194,
  \dodoi{10.1051/0004-6361/202039807}

\bibitem[{{Garrison-Kimmel} {et~al.}(2014){Garrison-Kimmel}, {Boylan-Kolchin},
  {Bullock}, \& {Lee}}]{elvis}
{Garrison-Kimmel}, S., {Boylan-Kolchin}, M., {Bullock}, J.~S., \& {Lee}, K.
  2014, \mnras, 438, 2578, \dodoi{10.1093/mnras/stt2377}

\bibitem[{{Genzel} {et~al.}(2006){Genzel}, {Tacconi}, {Eisenhauer},
  {F{\"o}rster Schreiber}, {Cimatti}, {Daddi}, {Bouch{\'e}}, {Davies},
  {Lehnert}, {Lutz}, {Nesvadba}, {Verma}, {Abuter}, {Shapiro}, {Sternberg},
  {Renzini}, {Kong}, {Arimoto}, \& {Mignoli}}]{genzel06}
{Genzel}, R., {Tacconi}, L.~J., {Eisenhauer}, F., {et~al.} 2006, \nat, 442,
  786, \dodoi{10.1038/nature05052}

\bibitem[{{Genzel} {et~al.}(2008){Genzel}, {Burkert}, {Bouch{\'e}}, {Cresci},
  {F{\"o}rster Schreiber}, {Shapley}, {Shapiro}, {Tacconi}, {Buschkamp},
  {Cimatti}, {Daddi}, {Davies}, {Eisenhauer}, {Erb}, {Genel}, {Gerhard},
  {Hicks}, {Lutz}, {Naab}, {Ott}, {Rabien}, {Renzini}, {Steidel}, {Sternberg},
  \& {Lilly}}]{genzel08}
{Genzel}, R., {Burkert}, A., {Bouch{\'e}}, N., {et~al.} 2008, \apj, 687, 59,
  \dodoi{10.1086/591840}

\bibitem[{{Grand} {et~al.}(2016){Grand}, {Springel}, {G{\'o}mez}, {Marinacci},
  {Pakmor}, {Campbell}, \& {Jenkins}}]{grand16}
{Grand}, R. J.~J., {Springel}, V., {G{\'o}mez}, F.~A., {et~al.} 2016, \mnras,
  459, 199, \dodoi{10.1093/mnras/stw601}

\bibitem[{{Grand} {et~al.}(2017){Grand}, {G{\'o}mez}, {Marinacci}, {Pakmor},
  {Springel}, {Campbell}, {Frenk}, {Jenkins}, \& {White}}]{auriga}
{Grand}, R.~J.~J., {G{\'o}mez}, F.~A., {Marinacci}, F., {et~al.} 2017, \mnras,
  467, 179, \dodoi{10.1093/mnras/stx071}

\bibitem[{{Grand} {et~al.}(2018){Grand}, {Helly}, {Fattahi}, {Cautun}, {Cole},
  {Cooper}, {Deason}, {Frenk}, {G{\'o}mez}, {Hunt}, {Marinacci}, {Pakmor},
  {Simpson}, {Springel}, \& {Xu}}]{aurigaia}
{Grand}, R. J.~J., {Helly}, J., {Fattahi}, A., {et~al.} 2018, \mnras, 481,
  1726, \dodoi{10.1093/mnras/sty2403}

\bibitem[{{Guo} {et~al.}(2015){Guo}, {Ferguson}, {Bell}, {Koo}, {Conselice},
  {Giavalisco}, {Kassin}, {Lu}, {Lucas}, {Mandelker}, {McIntosh}, {Primack},
  {Ravindranath}, {Barro}, {Ceverino}, {Dekel}, {Faber}, {Fang}, {Koekemoer},
  {Noeske}, {Rafelski}, \& {Straughn}}]{guo15}
{Guo}, Y., {Ferguson}, H.~C., {Bell}, E.~F., {et~al.} 2015, \apj, 800, 39,
  \dodoi{10.1088/0004-637X/800/1/39}

\bibitem[{{Guo} {et~al.}(2018){Guo}, {Rafelski}, {Bell}, {Conselice}, {Dekel},
  {Faber}, {Giavalisco}, {Koekemoer}, {Koo}, {Lu}, {Mandelker}, {Primack},
  {Ceverino}, {de Mello}, {Ferguson}, {Hathi}, {Kocevski}, {Lucas},
  {P{\'e}rez-Gonz{\'a}lez}, {Ravindranath}, {Soto}, {Straughn}, \&
  {Wang}}]{guo18}
{Guo}, Y., {Rafelski}, M., {Bell}, E.~F., {et~al.} 2018, \apj, 853, 108,
  \dodoi{10.3847/1538-4357/aaa018}

\bibitem[{{Gurvich} {et~al.}(2023){Gurvich}, {Stern}, {Faucher-Gigu{\`e}re},
  {Hopkins}, {Wetzel}, {Moreno}, {Hayward}, {Richings}, \& {Hafen}}]{gurvich22}
{Gurvich}, A.~B., {Stern}, J., {Faucher-Gigu{\`e}re}, C.-A., {et~al.} 2023,
  \mnras, 519, 2598, \dodoi{10.1093/mnras/stac3712}

\bibitem[{{Hafen} {et~al.}(2022){Hafen}, {Stern}, {Bullock}, {Gurvich}, {Yu},
  {Faucher-Gigu{\`e}re}, {Fielding}, {Angl{\'e}s-Alc{\'a}zar}, {Quataert},
  {Wetzel}, {Starkenburg}, {Boylan-Kolchin}, {Moreno}, {Feldmann}, {El-Badry},
  {Chan}, {Trapp}, {Kere{\v{s}}}, \& {Hopkins}}]{hafen22}
{Hafen}, Z., {Stern}, J., {Bullock}, J., {et~al.} 2022, \mnras, 514, 5056,
  \dodoi{10.1093/mnras/stac1603}

\bibitem[{{Hammer} {et~al.}(2007){Hammer}, {Puech}, {Chemin}, {Flores}, \&
  {Lehnert}}]{hammer07}
{Hammer}, F., {Puech}, M., {Chemin}, L., {Flores}, H., \& {Lehnert}, M.~D.
  2007, \apj, 662, 322, \dodoi{10.1086/516727}

\bibitem[{{Herrera-Camus} {et~al.}(2022){Herrera-Camus}, {F{\"o}rster
  Schreiber}, {Price}, {{\"U}bler}, {Bolatto}, {Davies}, {Fisher}, {Genzel},
  {Lutz}, {Naab}, {Nestor}, {Shimizu}, {Sternberg}, {Tacconi}, \&
  {Tadaki}}]{herreracamus22}
{Herrera-Camus}, R., {F{\"o}rster Schreiber}, N.~M., {Price}, S.~H., {et~al.}
  2022, \aap, 665, L8, \dodoi{10.1051/0004-6361/202142562}

\bibitem[{{Hopkins} {et~al.}(2014){Hopkins}, {Kere{\v s}}, {O{\~n}orbe},
  {Faucher-Gigu{\`e}re}, {Quataert}, {Murray}, \& {Bullock}}]{fire}
{Hopkins}, P.~F., {Kere{\v s}}, D., {O{\~n}orbe}, J., {et~al.} 2014, \mnras,
  445, 581, \dodoi{10.1093/mnras/stu1738}

\bibitem[{{Hopkins} {et~al.}(2018){Hopkins}, {Wetzel}, {Kere{\v{s}}},
  {Faucher-Gigu{\`e}re}, {Quataert}, {Boylan-Kolchin}, {Murray}, {Hayward},
  {Garrison-Kimmel}, {Hummels}, {Feldmann}, {Torrey}, {Ma},
  {Angl{\'e}s-Alc{\'a}zar}, {Su}, {Orr}, {Schmitz}, {Escala}, {Sanderson},
  {Grudi{\'c}}, {Hafen}, {Kim}, {Fitts}, {Bullock}, {Wheeler}, {Chan},
  {Elbert}, \& {Narayanan}}]{fire2}
{Hopkins}, P.~F., {Wetzel}, A., {Kere{\v{s}}}, D., {et~al.} 2018, \mnras, 480,
  800, \dodoi{10.1093/mnras/sty1690}

\bibitem[{{Hopkins} {et~al.}(2023){Hopkins}, {Gurvich}, {Shen}, {Hafen},
  {Grudi{\'c}}, {Kurinchi-Vendhan}, {Hayward}, {Jiang}, {Orr}, {Wetzel},
  {Kere{\v{s}}}, {Stern}, {Faucher-Gigu{\`e}re}, {Bullock}, {Wheeler},
  {El-Badry}, {Loebman}, {Moreno}, {Boylan-Kolchin}, \&
  {Quataert}}]{hopkins23disk}
{Hopkins}, P.~F., {Gurvich}, A.~B., {Shen}, X., {et~al.} 2023, \mnras, 525,
  2241, \dodoi{10.1093/mnras/stad1902}

\bibitem[{Hunter(2007)}]{matplotlib}
Hunter, J.~D. 2007, CSE, 9, 90, \dodoi{10.1109/MCSE.2007.55}

\bibitem[{{Jacobs} {et~al.}(2023){Jacobs}, {Glazebrook}, {Calabr{\`o}}, {Treu},
  {Nannayakkara}, {Jones}, {Merlin}, {Abraham}, {Stevens}, {Vulcani}, {Yang},
  {Bonchi}, {Boyett}, {Brada{\v{c}}}, {Castellano}, {Fontana}, {Marchesini},
  {Malkan}, {Mason}, {Morishita}, {Paris}, {Santini}, {Trenti}, \&
  {Wang}}]{jacobs22}
{Jacobs}, C., {Glazebrook}, K., {Calabr{\`o}}, A., {et~al.} 2023, \apjl, 948,
  L13, \dodoi{10.3847/2041-8213/accd6d}

\bibitem[{Jones {et~al.}(2001-2016)Jones, Oliphant, Peterson, {et~al.}}]{scipy}
Jones, E., Oliphant, T., Peterson, P., {et~al.} 2001-2016, http://www.scipy.org

\bibitem[{{Kere{\v{s}}} {et~al.}(2005){Kere{\v{s}}}, {Katz}, {Weinberg}, \&
  {Dav{\'e}}}]{keres05}
{Kere{\v{s}}}, D., {Katz}, N., {Weinberg}, D.~H., \& {Dav{\'e}}, R. 2005,
  \mnras, 363, 2, \dodoi{10.1111/j.1365-2966.2005.09451.x}

\bibitem[{{Khoperskov} {et~al.}(2023){Khoperskov}, {Minchev}, {Libeskind},
  {Belokurov}, {Steinmetz}, {Gomez}, {Grand}, {Hoffman}, {Knebe}, {Sorce},
  {Spaare}, {Tempel}, \& {Vogelsberger}}]{khoperskov22c}
{Khoperskov}, S., {Minchev}, I., {Libeskind}, N., {et~al.} 2023, \aap, 677,
  A91, \dodoi{10.1051/0004-6361/202244234}

\bibitem[{{Kohandel} {et~al.}(2020){Kohandel}, {Pallottini}, {Ferrara},
  {Carniani}, {Gallerani}, {Vallini}, {Zanella}, \& {Behrens}}]{kohandel20}
{Kohandel}, M., {Pallottini}, A., {Ferrara}, A., {et~al.} 2020, \mnras, 499,
  1250, \dodoi{10.1093/mnras/staa2792}

\bibitem[{{Kregel} {et~al.}(2005){Kregel}, {van der Kruit}, \&
  {Freeman}}]{kregel05}
{Kregel}, M., {van der Kruit}, P.~C., \& {Freeman}, K.~C. 2005, \mnras, 358,
  503, \dodoi{10.1111/j.1365-2966.2005.08855.x}

\bibitem[{{Kretschmer} {et~al.}(2022){Kretschmer}, {Dekel}, \&
  {Teyssier}}]{kretschmer22}
{Kretschmer}, M., {Dekel}, A., \& {Teyssier}, R. 2022, \mnras, 510, 3266,
  \dodoi{10.1093/mnras/stab3648}

\bibitem[{{Lelli} {et~al.}(2021){Lelli}, {Di Teodoro}, {Fraternali}, {Man},
  {Zhang}, {De Breuck}, {Davis}, \& {Maiolino}}]{lelli21}
{Lelli}, F., {Di Teodoro}, E.~M., {Fraternali}, F., {et~al.} 2021, Science,
  371, 713, \dodoi{10.1126/science.abc1893}

\bibitem[{{Livermore} {et~al.}(2015){Livermore}, {Jones}, {Richard}, {Bower},
  {Swinbank}, {Yuan}, {Edge}, {Ellis}, {Kewley}, {Smail}, {Coppin}, \&
  {Ebeling}}]{livermore15}
{Livermore}, R.~C., {Jones}, T.~A., {Richard}, J., {et~al.} 2015, \mnras, 450,
  1812, \dodoi{10.1093/mnras/stv686}

\bibitem[{{Marinacci} {et~al.}(2018){Marinacci}, {Vogelsberger}, {Pakmor},
  {Torrey}, {Springel}, {Hernquist}, {Nelson}, {Weinberger}, {Pillepich},
  {Naiman}, \& {Genel}}]{marinacci18}
{Marinacci}, F., {Vogelsberger}, M., {Pakmor}, R., {et~al.} 2018, \mnras, 480,
  5113, \dodoi{10.1093/mnras/sty2206}

\bibitem[{{McCluskey} {et~al.}(2024){McCluskey}, {Wetzel}, {Loebman}, {Moreno},
  {Faucher-Gigu{\`e}re}, \& {Hopkins}}]{mccluskey23}
{McCluskey}, F., {Wetzel}, A., {Loebman}, S.~R., {et~al.} 2024, \mnras, 527,
  6926, \dodoi{10.1093/mnras/stad3547}

\bibitem[{{Mo} {et~al.}(1998){Mo}, {Mao}, \& {White}}]{mo-mao-white98}
{Mo}, H.~J., {Mao}, S., \& {White}, S. D.~M. 1998, \mnras, 295, 319,
  \dodoi{10.1046/j.1365-8711.1998.01227.x}

\bibitem[{{Myeong} {et~al.}(2022){Myeong}, {Belokurov}, {Aguado}, {Evans},
  {Caldwell}, \& {Bradley}}]{myeong22}
{Myeong}, G.~C., {Belokurov}, V., {Aguado}, D.~S., {et~al.} 2022, \apj, 938,
  21, \dodoi{10.3847/1538-4357/ac8d68}

\bibitem[{{Naidu} {et~al.}(2022){Naidu}, {Oesch}, {van Dokkum}, {Nelson},
  {Suess}, {Brammer}, {Whitaker}, {Illingworth}, {Bouwens}, {Tacchella},
  {Matthee}, {Allen}, {Bezanson}, {Conroy}, {Labbe}, {Leja}, {Leonova},
  {Magee}, {Price}, {Setton}, {Strait}, {Stefanon}, {Toft}, {Weaver}, \&
  {Weibel}}]{naidu22}
{Naidu}, R.~P., {Oesch}, P.~A., {van Dokkum}, P., {et~al.} 2022, \apjl, 940,
  L14, \dodoi{10.3847/2041-8213/ac9b22}

\bibitem[{{Naiman} {et~al.}(2018){Naiman}, {Pillepich}, {Springel},
  {Ramirez-Ruiz}, {Torrey}, {Vogelsberger}, {Pakmor}, {Nelson}, {Marinacci},
  {Hernquist}, {Weinberger}, \& {Genel}}]{naiman18}
{Naiman}, J.~P., {Pillepich}, A., {Springel}, V., {et~al.} 2018, \mnras, 477,
  1206, \dodoi{10.1093/mnras/sty618}

\bibitem[{{Neeleman} {et~al.}(2020){Neeleman}, {Prochaska}, {Kanekar}, \&
  {Rafelski}}]{neeleman20}
{Neeleman}, M., {Prochaska}, J.~X., {Kanekar}, N., \& {Rafelski}, M. 2020,
  \nat, 581, 269, \dodoi{10.1038/s41586-020-2276-y}

\bibitem[{{Nelson} {et~al.}(2018){Nelson}, {Pillepich}, {Springel},
  {Weinberger}, {Hernquist}, {Pakmor}, {Genel}, {Torrey}, {Vogelsberger},
  {Kauffmann}, {Marinacci}, \& {Naiman}}]{nelson18}
{Nelson}, D., {Pillepich}, A., {Springel}, V., {et~al.} 2018, \mnras, 475, 624,
  \dodoi{10.1093/mnras/stx3040}

\bibitem[{{Nelson} {et~al.}(2019{\natexlab{a}}){Nelson}, {Springel},
  {Pillepich}, {Rodriguez-Gomez}, {Torrey}, {Genel}, {Vogelsberger}, {Pakmor},
  {Marinacci}, {Weinberger}, {Kelley}, {Lovell}, {Diemer}, \&
  {Hernquist}}]{nelson19b}
{Nelson}, D., {Springel}, V., {Pillepich}, A., {et~al.} 2019{\natexlab{a}},
  Computational Astrophysics and Cosmology, 6, 2,
  \dodoi{10.1186/s40668-019-0028-x}

\bibitem[{{Nelson} {et~al.}(2019{\natexlab{b}}){Nelson}, {Pillepich},
  {Springel}, {Pakmor}, {Weinberger}, {Genel}, {Torrey}, {Vogelsberger},
  {Marinacci}, \& {Hernquist}}]{nelson19}
{Nelson}, D., {Pillepich}, A., {Springel}, V., {et~al.} 2019{\natexlab{b}},
  \mnras, 490, 3234, \dodoi{10.1093/mnras/stz2306}

\bibitem[{{Nelson} {et~al.}(2023){Nelson}, {Suess}, {Bezanson}, {Price}, {van
  Dokkum}, {Leja}, {Wang}, {Whitaker}, {Labb{\'e}}, {Barrufet}, {Brammer},
  {Eisenstein}, {Gibson}, {Hartley}, {Johnson}, {Heintz}, {Mathews}, {Miller},
  {Oesch}, {Sandles}, {Setton}, {Speagle}, {Tacchella}, {Tadaki}, {{\"U}bler},
  \& {Weaver}}]{nelson22}
{Nelson}, E.~J., {Suess}, K.~A., {Bezanson}, R., {et~al.} 2023, \apjl, 948,
  L18, \dodoi{10.3847/2041-8213/acc1e1}

\bibitem[{{Nordstr{\"o}m} {et~al.}(2004){Nordstr{\"o}m}, {Mayor}, {Andersen},
  {Holmberg}, {Pont}, {J{\o}rgensen}, {Olsen}, {Udry}, \&
  {Mowlavi}}]{nordstrom04}
{Nordstr{\"o}m}, B., {Mayor}, M., {Andersen}, J., {et~al.} 2004, \aap, 418,
  989, \dodoi{10.1051/0004-6361:20035959}

\bibitem[{{Oser} {et~al.}(2010){Oser}, {Ostriker}, {Naab}, {Johansson}, \&
  {Burkert}}]{oser10}
{Oser}, L., {Ostriker}, J.~P., {Naab}, T., {Johansson}, P.~H., \& {Burkert}, A.
  2010, \apj, 725, 2312, \dodoi{10.1088/0004-637X/725/2/2312}

\bibitem[{{Pakmor} {et~al.}(2011){Pakmor}, {Bauer}, \& {Springel}}]{pakmor11}
{Pakmor}, R., {Bauer}, A., \& {Springel}, V. 2011, \mnras, 418, 1392,
  \dodoi{10.1111/j.1365-2966.2011.19591.x}

\bibitem[{{Pakmor} \& {Springel}(2013)}]{pakmor13}
{Pakmor}, R., \& {Springel}, V. 2013, \mnras, 432, 176,
  \dodoi{10.1093/mnras/stt428}

\bibitem[{{Pensabene} {et~al.}(2020){Pensabene}, {Carniani}, {Perna}, {Cresci},
  {Decarli}, {Maiolino}, \& {Marconi}}]{pensabene20}
{Pensabene}, A., {Carniani}, S., {Perna}, M., {et~al.} 2020, \aap, 637, A84,
  \dodoi{10.1051/0004-6361/201936634}

\bibitem[{{Pillepich} {et~al.}(2018{\natexlab{a}}){Pillepich}, {Nelson},
  {Hernquist}, {Springel}, {Pakmor}, {Torrey}, {Weinberger}, {Genel}, {Naiman},
  {Marinacci}, \& {Vogelsberger}}]{pillepich18b}
{Pillepich}, A., {Nelson}, D., {Hernquist}, L., {et~al.} 2018{\natexlab{a}},
  \mnras, 475, 648, \dodoi{10.1093/mnras/stx3112}

\bibitem[{{Pillepich} {et~al.}(2018{\natexlab{b}}){Pillepich}, {Springel},
  {Nelson}, {Genel}, {Naiman}, {Pakmor}, {Hernquist}, {Torrey}, {Vogelsberger},
  {Weinberger}, \& {Marinacci}}]{pillepich18}
{Pillepich}, A., {Springel}, V., {Nelson}, D., {et~al.} 2018{\natexlab{b}},
  \mnras, 473, 4077, \dodoi{10.1093/mnras/stx2656}

\bibitem[{{Pillepich} {et~al.}(2019){Pillepich}, {Nelson}, {Springel},
  {Pakmor}, {Torrey}, {Weinberger}, {Vogelsberger}, {Marinacci}, {Genel}, {van
  der Wel}, \& {Hernquist}}]{pillepich19}
{Pillepich}, A., {Nelson}, D., {Springel}, V., {et~al.} 2019, \mnras, 490,
  3196, \dodoi{10.1093/mnras/stz2338}

\bibitem[{{Pillepich} {et~al.}(2023){Pillepich}, {Sotillo-Ramos}, {Ramesh},
  {Nelson}, {Engler}, {Rodriguez-Gomez}, {Fournier}, {Donnari}, {Springel}, \&
  {Hernquist}}]{pillepich23}
{Pillepich}, A., {Sotillo-Ramos}, D., {Ramesh}, R., {et~al.} 2023, arXiv
  e-prints, arXiv:2303.16217, \dodoi{10.48550/arXiv.2303.16217}

\bibitem[{{Planck Collaboration} {et~al.}(2016){Planck Collaboration}, {Ade},
  {Aghanim}, {Arnaud}, {Ashdown}, {Aumont}, {Baccigalupi}, {Banday},
  {Barreiro}, {Bartlett}, {Bartolo}, {Battaner}, {Battye}, {Benabed},
  {Beno{\^\i}t}, {Benoit-L{\'e}vy}, {Bernard}, {Bersanelli}, {Bielewicz},
  {Bock}, {Bonaldi}, {Bonavera}, {Bond}, {Borrill}, {Bouchet}, {Boulanger},
  {Bucher}, {Burigana}, {Butler}, {Calabrese}, {Cardoso}, {Catalano},
  {Challinor}, {Chamballu}, {Chary}, {Chiang}, {Chluba}, {Christensen},
  {Church}, {Clements}, {Colombi}, {Colombo}, {Combet}, {Coulais}, {Crill},
  {Curto}, {Cuttaia}, {Danese}, {Davies}, {Davis}, {de Bernardis}, {de Rosa},
  {de Zotti}, {Delabrouille}, {D{\'e}sert}, {Di Valentino}, {Dickinson},
  {Diego}, {Dolag}, {Dole}, {Donzelli}, {Dor{\'e}}, {Douspis}, {Ducout},
  {Dunkley}, {Dupac}, {Efstathiou}, {Elsner}, {En{\ss}lin}, {Eriksen},
  {Farhang}, {Fergusson}, {Finelli}, {Forni}, {Frailis}, {Fraisse},
  {Franceschi}, {Frejsel}, {Galeotta}, {Galli}, {Ganga}, {Gauthier}, {Gerbino},
  {Ghosh}, {Giard}, {Giraud-H{\'e}raud}, {Giusarma}, {Gjerl{\o}w},
  {Gonz{\'a}lez-Nuevo}, {G{\'o}rski}, {Gratton}, {Gregorio}, {Gruppuso},
  {Gudmundsson}, {Hamann}, {Hansen}, {Hanson}, {Harrison}, {Helou},
  {Henrot-Versill{\'e}}, {Hern{\'a}ndez-Monteagudo}, {Herranz}, {Hildebrandt},
  {Hivon}, {Hobson}, {Holmes}, {Hornstrup}, {Hovest}, {Huang}, {Huffenberger},
  {Hurier}, {Jaffe}, {Jaffe}, {Jones}, {Juvela}, {Keih{\"a}nen}, {Keskitalo},
  {Kisner}, {Kneissl}, {Knoche}, {Knox}, {Kunz}, {Kurki-Suonio}, {Lagache},
  {L{\"a}hteenm{\"a}ki}, {Lamarre}, {Lasenby}, {Lattanzi}, {Lawrence}, {Leahy},
  {Leonardi}, {Lesgourgues}, {Levrier}, {Lewis}, {Liguori}, {Lilje},
  {Linden-V{\o}rnle}, {L{\'o}pez-Caniego}, {Lubin}, {Mac{\'\i}as-P{\'e}rez},
  {Maggio}, {Maino}, {Mandolesi}, {Mangilli}, {Marchini}, {Maris}, {Martin},
  {Martinelli}, {Mart{\'\i}nez-Gonz{\'a}lez}, {Masi}, {Matarrese}, {McGehee},
  {Meinhold}, {Melchiorri}, {Melin}, {Mendes}, {Mennella}, {Migliaccio},
  {Millea}, {Mitra}, {Miville-Desch{\^e}nes}, {Moneti}, {Montier}, {Morgante},
  {Mortlock}, {Moss}, {Munshi}, {Murphy}, {Naselsky}, {Nati}, {Natoli},
  {Netterfield}, {N{\o}rgaard-Nielsen}, {Noviello}, {Novikov}, {Novikov},
  {Oxborrow}, {Paci}, {Pagano}, {Pajot}, {Paladini}, {Paoletti}, {Partridge},
  {Pasian}, {Patanchon}, {Pearson}, {Perdereau}, {Perotto}, {Perrotta},
  {Pettorino}, {Piacentini}, {Piat}, {Pierpaoli}, {Pietrobon}, {Plaszczynski},
  {Pointecouteau}, {Polenta}, {Popa}, {Pratt}, {Pr{\'e}zeau}, {Prunet},
  {Puget}, {Rachen}, {Reach}, {Rebolo}, {Reinecke}, {Remazeilles}, {Renault},
  {Renzi}, {Ristorcelli}, {Rocha}, {Rosset}, {Rossetti}, {Roudier},
  {Rouill{\'e} d'Orfeuil}, {Rowan-Robinson}, {Rubi{\~n}o-Mart{\'\i}n},
  {Rusholme}, {Said}, {Salvatelli}, {Salvati}, {Sandri}, {Santos},
  {Savelainen}, {Savini}, {Scott}, {Seiffert}, {Serra}, {Shellard}, {Spencer},
  {Spinelli}, {Stolyarov}, {Stompor}, {Sudiwala}, {Sunyaev}, {Sutton},
  {Suur-Uski}, {Sygnet}, {Tauber}, {Terenzi}, {Toffolatti}, {Tomasi},
  {Tristram}, {Trombetti}, {Tucci}, {Tuovinen}, {T{\"u}rler}, {Umana},
  {Valenziano}, {Valiviita}, {Van Tent}, {Vielva}, {Villa}, {Wade}, {Wandelt},
  {Wehus}, {White}, {White}, {Wilkinson}, {Yvon}, {Zacchei}, \&
  {Zonca}}]{planck-xiii}
{Planck Collaboration}, {Ade}, P.~A.~R., {Aghanim}, N., {et~al.} 2016, \aap,
  594, A13, \dodoi{10.1051/0004-6361/201525830}

\bibitem[{{Posses} {et~al.}(2023){Posses}, {Aravena}, {Gonz{\'a}lez-L{\'o}pez},
  {Assef}, {Lambert}, {Jones}, {Bouwens}, {Brisbin}, {D{\'\i}az-Santos},
  {Herrera-Camus}, {Ricci}, \& {Smit}}]{posses23}
{Posses}, A.~C., {Aravena}, M., {Gonz{\'a}lez-L{\'o}pez}, J., {et~al.} 2023,
  \aap, 669, A46, \dodoi{10.1051/0004-6361/202243399}

\bibitem[{{Quirk} {et~al.}(2022){Quirk}, {Guhathakurta}, {Gilbert}, {Chemin},
  {Dalcanton}, {Williams}, {Seth}, {Patel}, {Fung}, {Tangirala}, \&
  {Yusufali}}]{quirk22}
{Quirk}, A. C.~N., {Guhathakurta}, P., {Gilbert}, K.~M., {et~al.} 2022, \aj,
  163, 166, \dodoi{10.3847/1538-3881/ac5324}

\bibitem[{{Renaud} {et~al.}(2021){Renaud}, {Agertz}, {Read}, {Ryde},
  {Andersson}, {Bensby}, {Rey}, \& {Feuillet}}]{renaud21}
{Renaud}, F., {Agertz}, O., {Read}, J.~I., {et~al.} 2021, \mnras, 503, 5846,
  \dodoi{10.1093/mnras/stab250}

\bibitem[{{Rix} {et~al.}(2022){Rix}, {Chandra}, {Andrae}, {Price-Whelan},
  {Weinberg}, {Conroy}, {Fouesneau}, {Hogg}, {De Angeli}, {Naidu}, {Xiang}, \&
  {Ruz-Mieres}}]{rix22}
{Rix}, H.-W., {Chandra}, V., {Andrae}, R., {et~al.} 2022, \apj, 941, 45,
  \dodoi{10.3847/1538-4357/ac9e01}

\bibitem[{{Rizzo} {et~al.}(2021){Rizzo}, {Vegetti}, {Fraternali}, {Stacey}, \&
  {Powell}}]{rizzo21}
{Rizzo}, F., {Vegetti}, S., {Fraternali}, F., {Stacey}, H.~R., \& {Powell}, D.
  2021, \mnras, 507, 3952, \dodoi{10.1093/mnras/stab2295}

\bibitem[{{Rizzo} {et~al.}(2020){Rizzo}, {Vegetti}, {Powell}, {Fraternali},
  {McKean}, {Stacey}, \& {White}}]{rizzo20}
{Rizzo}, F., {Vegetti}, S., {Powell}, D., {et~al.} 2020, \nat, 584, 201,
  \dodoi{10.1038/s41586-020-2572-6}

\bibitem[{{Robertson} {et~al.}(2023){Robertson}, {Tacchella}, {Johnson},
  {Hausen}, {Alabi}, {Boyett}, {Bunker}, {Carniani}, {Egami}, {Eisenstein},
  {Hainline}, {Helton}, {Ji}, {Kumari}, {Lyu}, {Maiolino}, {Nelson}, {Rieke},
  {Shivaei}, {Sun}, {{\"U}bler}, {Williams}, {Willmer}, \&
  {Witstok}}]{robertson23}
{Robertson}, B.~E., {Tacchella}, S., {Johnson}, B.~D., {et~al.} 2023, \apjl,
  942, L42, \dodoi{10.3847/2041-8213/aca086}

\bibitem[{{Rodriguez-Gomez} {et~al.}(2016){Rodriguez-Gomez}, {Pillepich},
  {Sales}, {Genel}, {Vogelsberger}, {Zhu}, {Wellons}, {Nelson}, {Torrey},
  {Springel}, {Ma}, \& {Hernquist}}]{rodriguez-gomez16}
{Rodriguez-Gomez}, V., {Pillepich}, A., {Sales}, L.~V., {et~al.} 2016, \mnras,
  458, 2371, \dodoi{10.1093/mnras/stw456}

\bibitem[{{Roman-Oliveira} {et~al.}(2023){Roman-Oliveira}, {Fraternali}, \&
  {Rizzo}}]{romanoliveira23}
{Roman-Oliveira}, F., {Fraternali}, F., \& {Rizzo}, F. 2023, \mnras, 521, 1045,
  \dodoi{10.1093/mnras/stad530}

\bibitem[{{Santistevan} {et~al.}(2020){Santistevan}, {Wetzel}, {El-Badry},
  {Bland-Hawthorn}, {Boylan-Kolchin}, {Bailin}, {Faucher-Gigu{\`e}re}, \&
  {Benincasa}}]{santistevan20}
{Santistevan}, I.~B., {Wetzel}, A., {El-Badry}, K., {et~al.} 2020, \mnras, 497,
  747, \dodoi{10.1093/mnras/staa1923}

\bibitem[{{Semenov} {et~al.}(2023){Semenov}, {Conroy}, {Chandra}, {Hernquist},
  \& {Nelson}}]{semenov23b}
{Semenov}, V.~A., {Conroy}, C., {Chandra}, V., {Hernquist}, L., \& {Nelson}, D.
  2023, arXiv e-prints, arXiv:2306.13125, \dodoi{10.48550/arXiv.2306.13125}

\bibitem[{{Simons} {et~al.}(2017){Simons}, {Kassin}, {Weiner}, {Faber},
  {Trump}, {Heckman}, {Koo}, {Pacifici}, {Primack}, {Snyder}, \& {de la
  Vega}}]{simons17}
{Simons}, R.~C., {Kassin}, S.~A., {Weiner}, B.~J., {et~al.} 2017, \apj, 843,
  46, \dodoi{10.3847/1538-4357/aa740c}

\bibitem[{{Smit} {et~al.}(2018){Smit}, {Bouwens}, {Carniani}, {Oesch},
  {Labb{\'e}}, {Illingworth}, {van der Werf}, {Bradley}, {Gonzalez}, {Hodge},
  {Holwerda}, {Maiolino}, \& {Zheng}}]{smit18}
{Smit}, R., {Bouwens}, R.~J., {Carniani}, S., {et~al.} 2018, \nat, 553, 178,
  \dodoi{10.1038/nature24631}

\bibitem[{{Somerville} \& {Dav{\'e}}(2015)}]{somerville-dave15}
{Somerville}, R.~S., \& {Dav{\'e}}, R. 2015, \araa, 53, 51,
  \dodoi{10.1146/annurev-astro-082812-140951}

\bibitem[{{Sotillo-Ramos} {et~al.}(2022){Sotillo-Ramos}, {Pillepich},
  {Donnari}, {Nelson}, {Eisert}, {Rodriguez-Gomez}, {Joshi}, {Vogelsberger}, \&
  {Hernquist}}]{sotillo-ramos22}
{Sotillo-Ramos}, D., {Pillepich}, A., {Donnari}, M., {et~al.} 2022, \mnras,
  516, 5404, \dodoi{10.1093/mnras/stac2586}

\bibitem[{{Springel}(2010)}]{arepo}
{Springel}, V. 2010, \mnras, 401, 791, \dodoi{10.1111/j.1365-2966.2009.15715.x}

\bibitem[{{Springel} \& {Hernquist}(2003)}]{sh03}
{Springel}, V., \& {Hernquist}, L. 2003, \mnras, 339, 289,
  \dodoi{10.1046/j.1365-8711.2003.06206.x}

\bibitem[{{Springel} {et~al.}(2018){Springel}, {Pakmor}, {Pillepich},
  {Weinberger}, {Nelson}, {Hernquist}, {Vogelsberger}, {Genel}, {Torrey},
  {Marinacci}, \& {Naiman}}]{springel18}
{Springel}, V., {Pakmor}, R., {Pillepich}, A., {et~al.} 2018, \mnras, 475, 676,
  \dodoi{10.1093/mnras/stx3304}

\bibitem[{{Stern} {et~al.}(2019){Stern}, {Fielding}, {Faucher-Gigu{\`e}re}, \&
  {Quataert}}]{stern19}
{Stern}, J., {Fielding}, D., {Faucher-Gigu{\`e}re}, C.-A., \& {Quataert}, E.
  2019, \mnras, 488, 2549, \dodoi{10.1093/mnras/stz1859}

\bibitem[{{Stern} {et~al.}(2020){Stern}, {Fielding}, {Faucher-Gigu{\`e}re}, \&
  {Quataert}}]{stern20}
---. 2020, \mnras, 492, 6042, \dodoi{10.1093/mnras/staa198}

\bibitem[{{Stern} {et~al.}(2023){Stern}, {Fielding}, {Hafen}, {Su}, {Naor},
  {Faucher-Gigu{\`e}re}, {Quataert}, \& {Bullock}}]{stern23}
{Stern}, J., {Fielding}, D., {Hafen}, Z., {et~al.} 2023, arXiv e-prints,
  arXiv:2306.00092, \dodoi{10.48550/arXiv.2306.00092}

\bibitem[{{Stern} {et~al.}(2021){Stern}, {Faucher-Gigu{\`e}re}, {Fielding},
  {Quataert}, {Hafen}, {Gurvich}, {Ma}, {Byrne}, {El-Badry},
  {Angl{\'e}s-Alc{\'a}zar}, {Chan}, {Feldmann}, {Kere{\v{s}}}, {Wetzel},
  {Murray}, \& {Hopkins}}]{stern21}
{Stern}, J., {Faucher-Gigu{\`e}re}, C.-A., {Fielding}, D., {et~al.} 2021, \apj,
  911, 88, \dodoi{10.3847/1538-4357/abd776}

\bibitem[{{Tacchella} {et~al.}(2019){Tacchella}, {Diemer}, {Hernquist},
  {Genel}, {Marinacci}, {Nelson}, {Pillepich}, {Rodriguez-Gomez}, {Sales},
  {Springel}, \& {Vogelsberger}}]{tacchella19}
{Tacchella}, S., {Diemer}, B., {Hernquist}, L., {et~al.} 2019, \mnras, 487,
  5416, \dodoi{10.1093/mnras/stz1657}

\bibitem[{{Tamfal} {et~al.}(2022){Tamfal}, {Mayer}, {Quinn}, {Babul}, {Madau},
  {Capelo}, {Shen}, \& {Staub}}]{tamfal22}
{Tamfal}, T., {Mayer}, L., {Quinn}, T.~R., {et~al.} 2022, \apj, 928, 106,
  \dodoi{10.3847/1538-4357/ac558e}

\bibitem[{{Toth} \& {Ostriker}(1992)}]{toth-ostriker92}
{Toth}, G., \& {Ostriker}, J.~P. 1992, \apj, 389, 5, \dodoi{10.1086/171185}

\bibitem[{{Tsukui} \& {Iguchi}(2021)}]{tsukui-iguchi21}
{Tsukui}, T., \& {Iguchi}, S. 2021, Science, 372, 1201,
  \dodoi{10.1126/science.abe9680}

\bibitem[{{van der Walt} {et~al.}(2011){van der Walt}, {Colbert}, \&
  {Varoquaux}}]{numpy_ndarray}
{van der Walt}, S., {Colbert}, S.~C., \& {Varoquaux}, G. 2011, CSE, 13, 22,
  \dodoi{10.1109/MCSE.2011.37}

\bibitem[{{Vasiliev} {et~al.}(2021){Vasiliev}, {Belokurov}, \&
  {Erkal}}]{vasiliev21}
{Vasiliev}, E., {Belokurov}, V., \& {Erkal}, D. 2021, \mnras, 501, 2279,
  \dodoi{10.1093/mnras/staa3673}

\bibitem[{{Vogelsberger} {et~al.}(2013){Vogelsberger}, {Genel}, {Sijacki},
  {Torrey}, {Springel}, \& {Hernquist}}]{vogelsberger13}
{Vogelsberger}, M., {Genel}, S., {Sijacki}, D., {et~al.} 2013, \mnras, 436,
  3031, \dodoi{10.1093/mnras/stt1789}

\bibitem[{{Watkins} {et~al.}(2010){Watkins}, {Evans}, \& {An}}]{watkins10}
{Watkins}, L.~L., {Evans}, N.~W., \& {An}, J.~H. 2010, \mnras, 406, 264,
  \dodoi{10.1111/j.1365-2966.2010.16708.x}

\bibitem[{{Weinberger} {et~al.}(2017){Weinberger}, {Springel}, {Hernquist},
  {Pillepich}, {Marinacci}, {Pakmor}, {Nelson}, {Genel}, {Vogelsberger},
  {Naiman}, \& {Torrey}}]{weinberger17}
{Weinberger}, R., {Springel}, V., {Hernquist}, L., {et~al.} 2017, \mnras, 465,
  3291, \dodoi{10.1093/mnras/stw2944}

\bibitem[{{Wetzel} {et~al.}(2023){Wetzel}, {Hayward}, {Sanderson}, {Ma},
  {Angl{\'e}s-Alc{\'a}zar}, {Feldmann}, {Chan}, {El-Badry}, {Wheeler},
  {Garrison-Kimmel}, {Nikakhtar}, {Panithanpaisal}, {Arora}, {Gurvich},
  {Samuel}, {Sameie}, {Pandya}, {Hafen}, {Hummels}, {Loebman},
  {Boylan-Kolchin}, {Bullock}, {Faucher-Gigu{\`e}re}, {Kere{\v{s}}},
  {Quataert}, \& {Hopkins}}]{fire2-release}
{Wetzel}, A., {Hayward}, C.~C., {Sanderson}, R.~E., {et~al.} 2023, \apjs, 265,
  44, \dodoi{10.3847/1538-4365/acb99a}

\bibitem[{{Wisnioski} {et~al.}(2015){Wisnioski}, {F{\"o}rster Schreiber},
  {Wuyts}, {Wuyts}, {Bandara}, {Wilman}, {Genzel}, {Bender}, {Davies},
  {Fossati}, {Lang}, {Mendel}, {Beifiori}, {Brammer}, {Chan}, {Fabricius},
  {Fudamoto}, {Kulkarni}, {Kurk}, {Lutz}, {Nelson}, {Momcheva}, {Rosario},
  {Saglia}, {Seitz}, {Tacconi}, \& {van Dokkum}}]{wisnioski15}
{Wisnioski}, E., {F{\"o}rster Schreiber}, N.~M., {Wuyts}, S., {et~al.} 2015,
  \apj, 799, 209, \dodoi{10.1088/0004-637X/799/2/209}

\bibitem[{{Wyse}(2001)}]{wyse01}
{Wyse}, R.~F.~G. 2001, in Astronomical Society of the Pacific Conference
  Series, Vol. 230, Galaxy Disks and Disk Galaxies, ed. J.~G. {Funes} \& E.~M.
  {Corsini}, 71--80, \dodoi{10.48550/arXiv.astro-ph/0012270}

\bibitem[{{Xiang} \& {Rix}(2022)}]{xiang-rix22}
{Xiang}, M., \& {Rix}, H.-W. 2022, \nat, 603, 599,
  \dodoi{10.1038/s41586-022-04496-5}

\bibitem[{{Yu} {et~al.}(2023){Yu}, {Bullock}, {Gurvich}, {Hafen}, {Stern},
  {Boylan-Kolchin}, {Faucher-Gigu{\`e}re}, {Wetzel}, {Hopkins}, \&
  {Moreno}}]{yu22}
{Yu}, S., {Bullock}, J.~S., {Gurvich}, A.~B., {et~al.} 2023, \mnras, 523, 6220,
  \dodoi{10.1093/mnras/stad1806}

\bibitem[{{Zaritsky} {et~al.}(2020){Zaritsky}, {Conroy}, {Zhang}, {Naidu},
  {Bonaca}, {Caldwell}, {Cargile}, \& {Johnson}}]{zaritsky20}
{Zaritsky}, D., {Conroy}, C., {Zhang}, H., {et~al.} 2020, \apj, 888, 114,
  \dodoi{10.3847/1538-4357/ab5b93}

\end{thebibliography}

\end{document}